\documentclass[acmsmall]{acmart}

\usepackage{array}
\usepackage{multirow}
\usepackage{booktabs}
\usepackage{threeparttable}
\usepackage{subfigure}

\usepackage{color}
\usepackage{colortbl}
\definecolor{tabcolor}{rgb}{1.0,0,0}
\AtBeginDocument{%
  }

\setcopyright{acmcopyright}
\acmDOI{XXXXXXX.XXXXXXX}

\begin{document}

\title{Deep Network for Image Compressed Sensing Coding Using Local Structural Sampling}

\author{Wenxue Cui}
\email{wxcui@hit.edu.cn}
\orcid{0000-0001-8656-0954}
\author{Xingtao Wang}
\email{xtwang@hit.edu.cn}
\orcid{0000-0002-5763-2493}
\affiliation{%
  \institution{Department of Computer Science and Technology, Harbin Institute of Technology}
  \streetaddress{West Dazhi Street}
  \city{Harbin}
  \state{Heilongjiang}
  \country{China}
  \postcode{150001}
}

\author{Xiaopeng Fan}
\email{fxp@hit.edu.cn}
\orcid{0000-0002-9660-3636}
\author{Shaohui Liu}
\email{shliu@hit.edu.cn}
\orcid{0000-0002-1810-5412}
\affiliation{%
  \institution{Department of Computer Science and Technology, Harbin Institute of Technology}
  \streetaddress{West Dazhi Street}
  \city{Harbin}
  \state{Heilongjiang}
  \country{China}
  \postcode{150001}
}
\affiliation{%
  \institution{Peng Cheng Laboratory}
  \city{Shenzhen}
  \state{Guangdong}
  \country{China}
}

\author{Xinwei Gao}
\email{vitogao@tencent.com}
\orcid{0009-0008-7399-601X}
\affiliation{%
  \institution{Wechat Business Group, Tencent}
  \city{Shenzhen}
  \state{Guangdong}
  \country{China}
  \postcode{518057}
}

\author{Debin Zhao}
\authornote{Corresponding Author.}
\email{dbzhao@hit.edu.cn}
\orcid{0000-0003-3434-9967}
\affiliation{%
  \institution{Department of Computer Science and Technology, Harbin Institute of Technology}
  \streetaddress{West Dazhi Street}
  \city{Harbin}
  \state{Heilongjiang}
  \country{China}
  \postcode{150001}
}
\affiliation{%
  \institution{Peng Cheng Laboratory}
  \city{Shenzhen}
  \state{Guangdong}
  \country{China}
}

\renewcommand{\shortauthors}{Wenxue Cui et al.}

\begin{abstract}
  Existing image compressed sensing (CS) coding frameworks usually solve an inverse problem based on measurement coding and optimization-based image reconstruction, which still exist the following two challenges: 1) The widely used random sampling matrix, such as the Gaussian Random Matrix (GRM), usually leads to low measurement coding efficiency. 2) The optimization-based reconstruction methods generally maintain a much higher computational complexity. In this paper, we propose a new CNN based image CS coding framework using local structural sampling (dubbed CSCNet) that includes three functional modules: local structural sampling, measurement coding and Laplacian pyramid reconstruction. In the proposed framework, instead of GRM, a new local structural sampling matrix is first developed, which is able to enhance the correlation between the measurements through a local perceptual sampling strategy. Besides, the designed local structural sampling matrix can be jointly optimized with the other functional modules during training process. After sampling, the measurements with high correlations are produced, which are then coded into final bitstreams by the third-party image codec. At last, a Laplacian pyramid reconstruction network is proposed to efficiently recover the target image from the measurement domain to the image domain. Extensive experimental results demonstrate that the proposed scheme outperforms the existing state-of-the-art CS coding methods, while maintaining fast computational speed.
\end{abstract}




\maketitle

\section{Introduction}
According to the Nyquist-Shannon sampling theorem, the traditional image acquisition system usually acquires a set of highly dense samples at a sampling rate not less than twice the highest frequency of the signal, and then compresses the signal to remove redundancy by a heavy-duty compressor for storage and transmission. However, this traditional image acquisition system may not be suitable for the resource-deficient visual communications, such as inexpensive resource-deprived sensors and computing-limited processors. Besides, in the medical imaging applications, it is important to reduce the time of the patients' exposure in the electromagnetic radiation. The emerging technology of Compressed Sensing~\cite{donoho2006compressed, candes2008introduction} (CS) leads to a new paradigm for image acquisition that performs sampling and compression jointly with much lower encoding complexity. Specifically, the CS theory shows that if a signal is sparse in a certain domain, it can be accurately recovered from a small number of its linear measurements much less than that determined by the Nyquist sampling theorem. The possible reduction of sampling rate is attractive for diverse imaging applications such as wireless sensor network~\cite{wu2016privacy,10.1145/2818712}, Magnetic Resonance Imaging~\cite{Lustig2008Compressed,10032194} (MRI), data encryption~\cite{li2016image,8902011} and Compressive Imaging~\cite{tang2019feature,liu2022deep}.

In the study of CS, the two main challenges are the design of sampling matrix~\cite{chen2022content,yan2014shrinkage} and the development of reconstruction solvers~\cite{10.1145/3447431}. Recently, some optimization-based schemes~\cite{zhang2014group,dong2014compressive,zhao2016nonconvex} and deep network-based CS methods~\cite{shi2019image,9199540,9019857} are proposed to deal with these two challenges and achieve great success. However, the signal recovery from the continuous measurement domain may not be favored in the signal storage and transmission. Therefore, the signal acquisition is usually performed by Analog-to-Digital Converters (ADC) that bounds each measurement into a predefined value with a finite number of bits. More importantly, with the increasing applications of CS in recent years, a large number of CS signals are generated, which need to be efficiently stored and transmitted. As a result, the study of CS coding schemes attracts much attentions.

For CS sampling, the Block-based CS~\cite{gan2007block} (BCS) sampling mechanism is usually adopted, in which the measurement is produced in a block-by-block sampling manner. Based on BCS, a variety of sampling matrices, such as the Gaussian Random Matrix (GRM)~\cite{Mun2012DPCM,zhang2012compressed} and the structure-based sampling matrix~\cite{gao2015block,6738003} are developed. However, these sampling matrices are all signal independent, which usually ignore the characteristics of the input signal, thus leading to unsatisfactory reconstruction quality. To release the above problem, some deep network-based sampling matrices are proposed in~\cite{shi2019image,9298950}, which can be jointly optimized with the reconstruction module during training process and achieve better reconstruction performance. However, these deep network-based sampling matrices usually focus only on the reconstruction quality, and do not consider the measurement coding efficiency, which generally bring about inefficient measurement coding performance, limiting the efficient storage and transmission of measured signals.

For CS measurement coding, a variety of CS coding methods are proposed. In the early stage of CS coding researching, quantifying the measurement directly~\cite{Yang2013Variational,Jacques2011Dequantizing} is widely studied, which usually causes lower rate-distortion performance. To mitigate this problem, some prediction based CS coding frameworks are proposed in~\cite{Mun2012DPCM, Zhang2013Spatially,gao2015block,9287074,9508849} to make use of the correlations between the measurements. More specifically, these prediction-based CS coding methods usually first predict the current measurements by the other observations sampled from the neighboring image blocks, and then further encode the measurement residuals into bitstreams. Compared with the methods of directly quantifying the measurements, the prediction-based CS coding algorithms achieve better coding performance. However, there is still a large gap compared to the traditional third-party image codecs, such as JPEG2000 and BPG (based on HEVC-Intra). In addition, the optimization-based solvers are usually used in these CS coding methods to reconstruct the target image in an iterative manner, which increase the computational complexity, thus affecting the execution efficiency and limiting the practical applications significantly.

Compared with the traditional image codecs, the aforementioned image CS coding methods can be directly used in CS applications. For example, these image CS coding schemes can be integrated into the single-pixel camera~\cite{4472247} or the lensless camera~\cite{6738433}. For simplicity, we abbreviate the above CS coding methods as CSC. In fact, in addition to CSC, there is also another kind of CS-based image coding method~\cite{Liu2016Compressive,chen2019compressive,8110646}, which usually takes CS as a dimensionality reduction tool to compress images and we abbreviate this kind of CS-based coding algorithm as CSBC. It is noted that since sampling ratio is an important factor in most CS-related application systems, the coding efficiency of CSC methods is usually analyzed under different sampling ratios. In other words, the CSC methods usually pay more attention to the tunability of sampling ratio, which ensures adaptation with most CS systems. Instead, the CSBC methods generally focus only on coding efficiency and does not pay much attention to the controllability of sampling ratio. 

In this paper, we mainly focus on the CSC method and propose a new Compressed Sensing Coding Network (CSCNet) using local structural sampling for image CS coding. In the proposed framework, a well-designed local structural sampling matrix is first developed to locally perceive the input image. It is noted that the developed sampling matrix is highly sparse because of its locally structured design, which not only can be easily implemented in hardware for compressive imaging, but also can be jointly optimized with the other functional modules during training process. After sampling, the measurements with high correlations are produced, which are then coded into final bitstreams by a certain existing third-party image codec. At last, a convolutional Laplacian pyramid architecture is proposed to reconstruct the target image from the measurement domain to the image domain. It should be noted that the proposed framework can be trained in an end-to-end manner, which facilitates the communication among different functional modules for better coding performance. Experimental results manifest that the proposed CSCNet outperforms the other state-of-the-art CS coding methods, while maintaining fast computational speed.

The main contributions are summarized as follows:

\textbf{1)} A new CNN based image CS coding framework using local structural sampling is proposed, which includes three functional modules: local structural sampling, measurement coding and Laplacian pyramid reconstruction.

\textbf{2)} A learnable local structural sampling matrix with high sparsity is designed, which not only can be easily implemented in hardware for compressive imaging because of its highly sparse characteristic, but also can be applied to generate highly correlated measurements for efficient measurement coding.

\textbf{3)} A convolutional Laplacian pyramid network is developed to progressively reconstruct the target image from measurement domain to the reconstructed image domain.

A preliminary version of this work was presented earlier in~\cite{Cui2018An}. This work improves the preliminary version in the following three aspects. First, instead of using the learned GRM in~\cite{Cui2018An}, a learnable local structural sampling matrix is developed, which can produce highly correlated measurements during the sampling process, thus enhancing the prediction accuracy among different sets of measurements and improving coding efficiency. Second, the produced measurements are directly coded with the existing third-party image codec, instead of the arithmetic coding in~\cite{8962013}, to generate the final bitstreams.
Third, a convolutional Laplacian pyramid network is designed to reconstruct the target image progressively from the quantized measurement domain to the reconstructed image domain.

The remainder of this paper is organized as follows: Section~\ref{section:a2} reviews the related works. Section~\ref{section:a3} elaborates more details of the proposed compressed sensing coding framework. Specifically, the local structural sampling is demonstrated in Subsection~\ref{section:a31}. The measurement coding is presented in Subsection~\ref{section:a32}. Subsection~\ref{section:a33} provides the architecture details of the proposed Laplacian pyramid reconstruction network. Section~\ref{section:a4} provides more implementation details and experimental results compared with CS coding methods (CSC), CS-based image coding algorithms (CSBC) and existing third-party image coding standards. Section~\ref{section:a6} concludes the paper.

\begin{figure*}
\centering
\includegraphics[width=\textwidth]{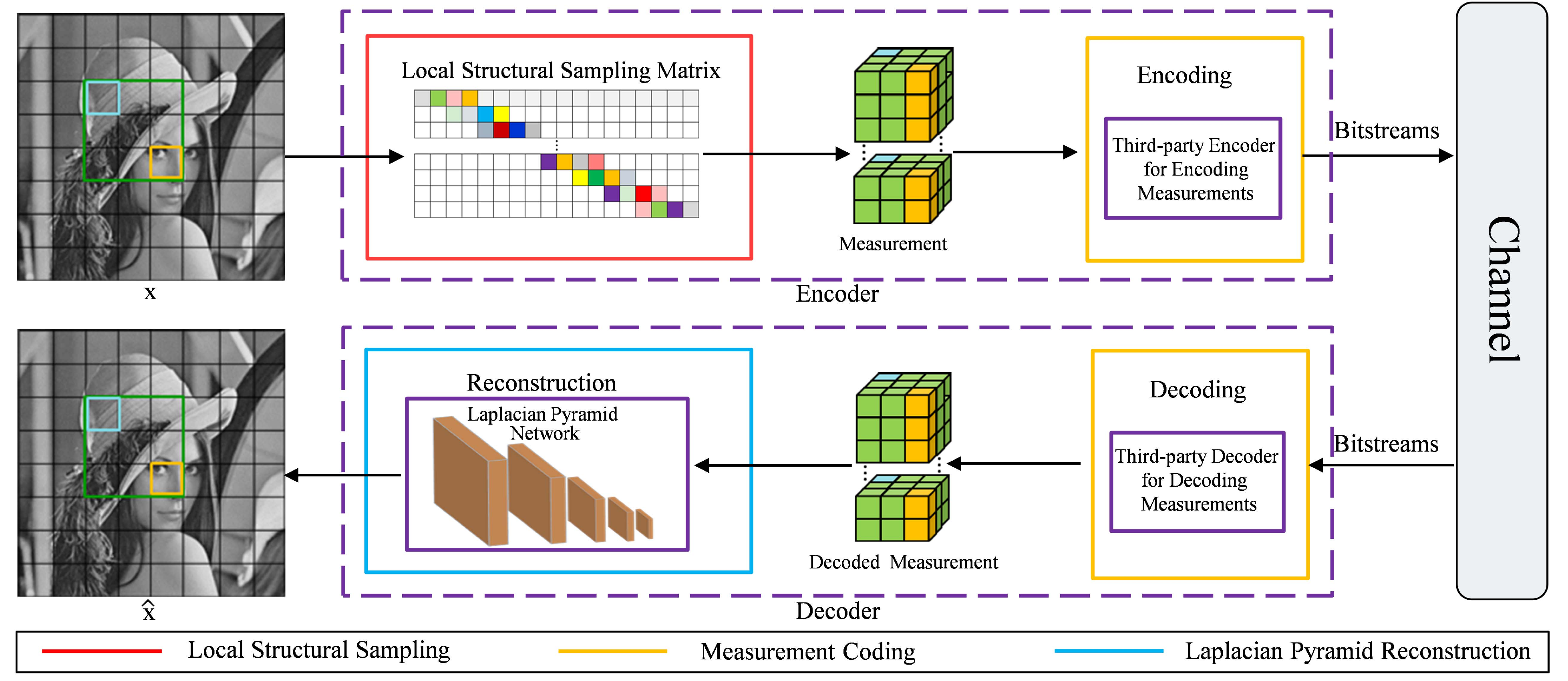}
\vspace{-0.25in}\caption{Diagram of the proposed image CS coding framework CSCNet. Three functional modules are included: local structural sampling, measurement coding and Laplacian pyramid reconstruction.}
\vspace{-0.16in}
\label{Fig:framework}
\end{figure*}

\section{Background and Related Works}
\label{section:a2}

\subsection{Image Compressed Sensing}
Compressed sensing (CS) has drawn quite an amount of attention as a joint sampling and compression methodology. The CS theory~\cite{candes2008introduction} shows that if a signal $x\in \mathbb{R}^{N}$ is sparse in a certain domain $\Psi$, it can be recovered with high probability from a small number of its linear measurements $y\in \mathbb{R}^{M}$. Mathematically, the CS sampling process can be performed by the following linear transform
\vspace{-0.01in}
\begin{equation}
y = \Phi x
\vspace{-0.01in}
\end{equation}
where $\Phi\in \mathbb{R}^{M\times N}$ is the sampling matrix. CS aims to recover the signal $x$ from its linear measurements $y$ efficiently. Because $M\ll N$, this inverse problem is ill-posed. Recently, many CS reconstruction methods are proposed, which can be roughly grouped into the following two categories: optimization-based methods and deep network-based methods.

Given the linear measurements $y$, traditional optimization-based image CS methods usually reconstruct the original image $x$ by solving an optimization problem:
\begin{equation}
\hat{x}=\mathop{\arg}\mathop{\min}_{x}\frac{1}{2} \| \Phi x-y \|_{2}^{2}+\lambda \| \Psi x \|_{p}
\label{Eq:a2}
\end{equation}
where $\Psi x$ denotes the sparse coefficients with respect to the transform $\Psi$ and the sparsity is characterized by the $p$ norm. $\lambda$ is the regularization parameter to control the sparsity item. To solve Eq.~\ref{Eq:a2}, a lot of sparsity-regularized based methods have been proposed, such as the greedy algorithms~\cite{mallat1993matching,tropp2007signal} and the convex-optimization algorithms~\cite{daubechies2004iterative,wright2009sparse}. To enhance the reconstructed quality, some sophisticated models are established to explore more image priors~\cite{zhang2014group,li2013tval3,Metzler2016From,7505983}. Many of these approaches have led to significant improvements. However, these optimization-based algorithms generally have very high computational complexity because of their iterative solvers, thus limiting their practical applications.

In recent years, some deep network-based methods are developed for image CS reconstruction. Specifically, in~\cite{mousavi2015deep}, Mousavi $\emph{et al.}$ first propose a stacked denoising autoencoder (SDA) to capture statistical dependencies between the elements of the signal. However, the fully connected layer (FCN) utilized in SDA leads to a huge amount of learnable parameters. In order to relieve this problem, several CNN based reconstruction methods~\cite{Yao2017DR2,shi2019image} are proposed, which usually build a direct deep mapping from the measurement domain to the original image domain. Considering the black box characteristics of the above CS networks, some optimization inspired CS reconstruction networks are presented~\cite{zhang2018ista,9019857,9298950,song2021memory,10124848}, in which the neural networks are usually embedded into some optimization-based methods to enjoy better interpretability. More recently, several deep neural network-based scalable CS architectures are proposed in~\cite{xu2018lapran,shi2019scalable,9467810}, which achieve scalable sampling and reconstruction with only one model, thus enhancing the flexibility and practicability of CS greatly. In addition to the image signals, compressed sensing is also applied to the audio and video signals~\cite{xu2014compressive,9025255}. However, these aforementioned CS methods are not real compression tools in the strict information theoretic sense, but they can be only seen as the technologies of dimensionality reduction.

\begin{figure*}[t]
\vskip -0.02in
\centering
\includegraphics[width=10.8cm]{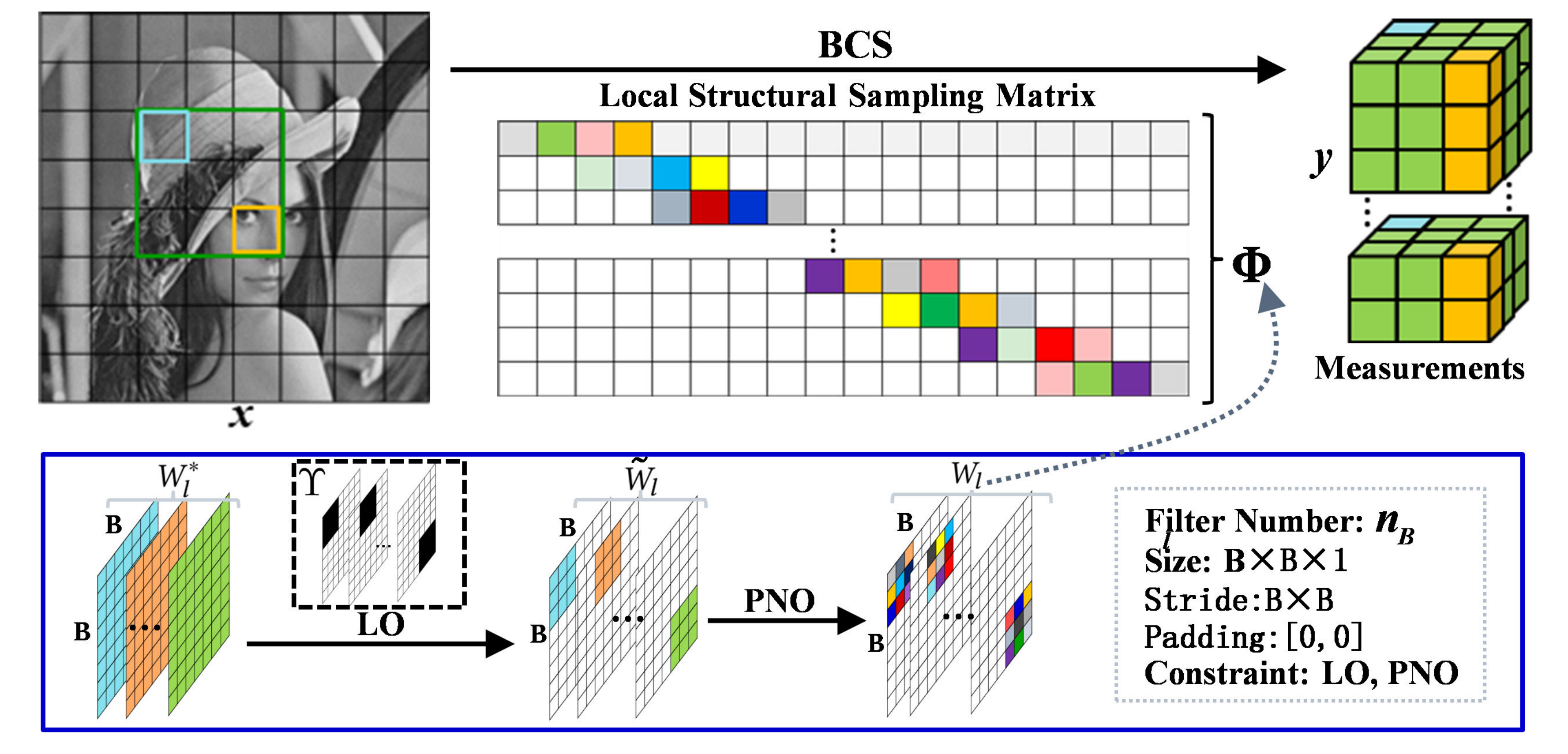}
\vspace{-0.1in} \caption{Diagram of the proposed local structural sampling. LO is the localization operation, and PNO is the positive normalization operation. The dashed box shows the configuration of our local sampling network.}
\vspace{-0.16in}
\label{Fig:local_sampling}
\end{figure*}

\subsection{Image Compressed Sensing Coding}

In order to facilitate the storage and transmission of the produced measurements, the quantization module is generally indispensable, in which the uniform scalar quantizer is usually utilized because of its simplicity and flexibility. However, quantifying the measurement directly~\cite{Yang2013Variational,Jacques2011Dequantizing} usually results in poor rate distortion performance. Inspired by the prediction techniques in recent video coding frameworks, such as H.264~\cite{Wiegand2003Overview} and HEVC~\cite{Sullivan2012Overview}, several prediction-based CS coding frameworks are presented. Specifically, Mun and Fowler~\cite{Mun2012DPCM} first propose a block-based quantized compressed sensing model of natural images with different pulse-code modulation (DPCM). Subsequently, Zhang $\emph{et al.}$~\cite{Zhang2013Spatially} extend the work in~\cite{Mun2012DPCM} by exploring more spatial correlation between the measurements of neighboring image blocks. In recent years, fueled by the powerful learning ability of deep networks, our early work~\cite{Cui2018An} constructs a CNN-based CS coding framework, which explores the learned GRM as sampling matrix, uses the arithmetic coding as entropy coder, and applies a convolutional network to reconstruct the target image. Subsequently, Yuan $\emph{et al}$~\cite{8962013} present an end-to-end image CS coding system, which integrates the conventional compressed sampling and reconstruction with quantization and entropy coding.

Recently, several locally perceived sampling matrices are proposed~\cite{gao2015block,9508849}, which are able to enhance the correlations between the measurements for efficient measurement coding. However, the elements of sampling matrices in these literatures are actually obtained in a random manner. In other words, the above random sampling matrices are all signal independent, which usually ignore the characteristics of the input signal, thus leading to limited reconstruction quality in most cases. Contrastively, in the proposed local structural sampling matrix, the elements can be automatically learned from a large amount of training data, which is able to facilitate the mining and utilization of prior knowledge in massive data, thus enhancing sampling efficiency. In addition to the sampling matrix, the manually designed measurement coding tools and the optimization-based reconstruction methods in~\cite{gao2015block,9508849} also limit their coding performance greatly.

Apparently, the above algorithms belong to the image CS coding (CSC) method. Compared with CSC, the CS-based image coding (CSBC) method is also widely studied in recent years. For example, in~\cite{Liu2016Compressive}, a CS-based image coding framework for resource-deficient visual communication is proposed, in which the third-party image codec is further utilized to compress the produced measurements. Recently, in order to efficiently establish the correlations between different sets of measurements, Chen $\emph{et al.}$~\cite{chen2019compressive} propose a multi-layer convolutional network, which explores the correlations between the measurements and recovers the target image progressively. It should be noted that the CSC method usually focuses on the compression performance under different sampling ratios, as sampling ratio is an important factor in most CS-related applications. While the CSBC method generally focuses only on coding efficiency and does not pay much attention to sampling ratio. Therefore, CSBC may not be suitable for MRI or other CS-related applications.

\begin{figure*}[t]
\vskip -0.02in
\centering
\includegraphics[width=13.8cm]{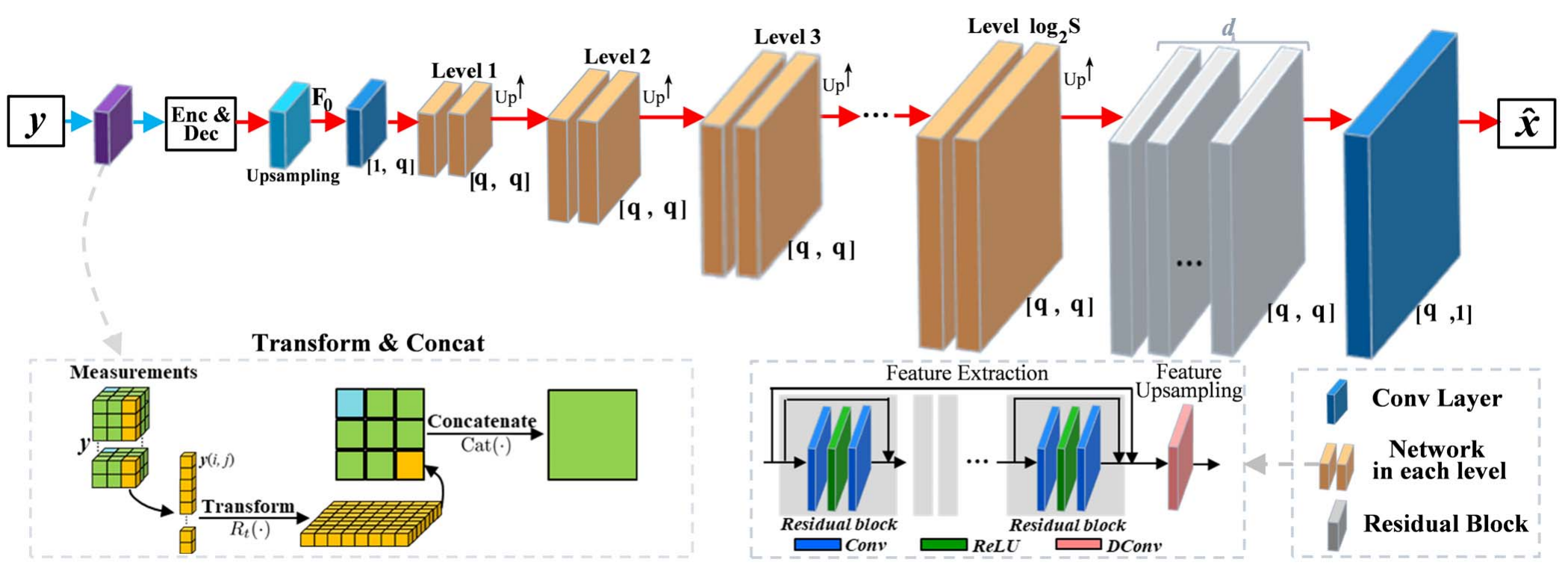}
\vspace{-0.1in} \caption{The details of the Laplacian pyramid reconstruction network (red arrow). The architectures in the dashed boxes show the details of ``Transform+Concat'' and the structure of each level respectively. The double values in brackets represent the channel number of input feature and output feature respectively. }
\vspace{-0.18in}
\label{Fig:reconstruction}
\end{figure*}

\subsection{Deep Neural Network-Based Image/Video Coding}

In recent years, inspired by the powerful learning ability of deep networks~\cite{10049122}, some deep network-based methods are proposed for image and video coding. Firstly, in order to obtain more efficient intermediate representations, several auto-encoder based coding frameworks are presented in~\cite{9050860,9359473,9067005}, which aim to build a deep mapping from the input image to the representation space in the encoder and reconstruct the original image in the decoder. Subsequently, several resampling based coding methods are developed in~\cite{7999241,8445655,8629276}, in which a third-party codec is usually used as an anchor to compress the compact representations. Later, the idea of resampling is successfully applied to the video coding standards, such as the works in~\cite{7982641,8554306,8476610}, which also obtains better performance. More recently, considering the problem of poor reconstruction quality in case of low bit rates, Generative Adversarial Network (GAN) is applied to the coding task~\cite{8574895,9010721}, which greatly improves the visual quality of the decoded image. By using the deep neural networks, the aforementioned coding methods achieve higher compression efficiency. However, these existing deep network-based coding methods take the original images or videos as input, which are captured by the traditional sampling system designed according to the Nyquist sampling theorem. Comparatively, the proposed method can be used to compress the measurements captured by the compressed sensing sampling system designed according to CS sampling theory.

\section{Image CS Coding Network Using Local Structural Sampling}

\label{section:a3}

Fig.~\ref{Fig:framework} shows the entire proposed image CS coding framework CSCNet, in which three functional modules are included: local structural sampling, measurement coding, and Laplacian pyramid reconstruction. It is worth noting that CNN is integrated into the proposed CSCNet. Specifically, in the image sampling module, a well-designed sampling network is first developed to learn a local structural sampling matrix for image sampling. After sampling process, the measurements with high correlations are produced, which are subsequently coded into bitstreams for measurement storage and transmission in the measurement coding module. Finally, a Laplacian pyramid reconstruction network is designed to learn an end-to-end mapping from the CS measurements to the target image.

\subsection{Local Structural Sampling}
\label{section:a31}
Local sampling aims to perceive the local structural information of the current image block, which is actually different from the global sampling (such as GRM) that perceives the global information of the given signal. Compared with the global sampling matrix, the local sampling matrix can effectively keep high correlations between the measurements, which usually brings about better coding performance~\cite{Liu2016Compressive}. Inspired by the local sampling method~\cite{gao2015block}, we propose a new learnable local structural sampling matrix, which can be jointly optimized with the other functional modules during training process.  Specifically, we use the convolutional neural network to learn the designed local structural sampling matrix, which will be described in detail below.

In block-based CS (BCS), the image $x$ with size $w \times h$ is first divided into non-overlapping blocks $x_{(i,j)}$ of size $B\times B$, where $i\in \{1,2,...,\frac{w}{B}\}$ and $j\in \{1,2,...,\frac{h}{B}\}$ are the position indexes of the current image block. Then, a sampling matrix $\Phi$ of size $n_{B}\times B^{2}$ is usually used to acquire the CS measurements. Specifically, given the sampling ratio (R) $\frac{M}{N}$, there are $n_{B}=\lfloor\frac{M}{N}B^{2}\rfloor$ rows in the sampling matrix $\Phi$ to obtain $n_{B}$ CS measurements for each image block. As above, the whole sampling process can be expressed as
\vspace{-0.01in}
\begin{equation}
y_{(i,j)}=\Phi x_{(i,j)}
\label{Eq:100}
\vspace{-0.01in}
\end{equation}
where $x_{(i,j)}$ is the current image block, and $y_{(i,j)}$ indicates the produced linear measurements of the current image block.

\begin{figure*}[t]
\centering
\hspace{-0.052in}
\subfigure{
\includegraphics[width=1.01in]{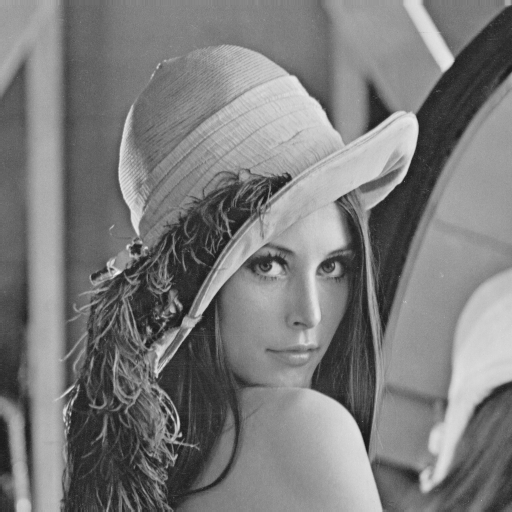}}
\hspace{0.02in}
\subfigure{
\includegraphics[width=1.01in]{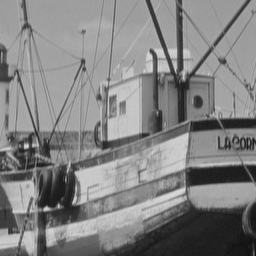}}
\hspace{0.02in}
\subfigure{
\includegraphics[width=1.01in]{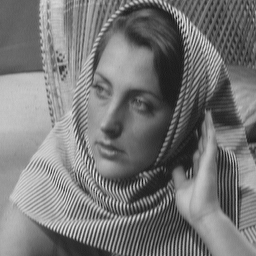}}
\hspace{0.02in}
\subfigure{
\includegraphics[width=1.01in]{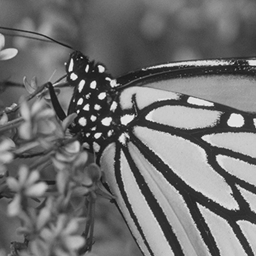}}
\vskip -0.07in \tiny{\quad Lenna \quad\quad\quad\quad\quad\quad\quad\quad\quad\quad\quad Boats \quad\quad\quad\quad\quad\quad\quad\quad\quad\quad Barbara \quad\quad\quad\quad\quad\quad\quad\quad\quad\quad Monarch}
\\
\vspace{-0.042in}
\hspace{-0.01in}
\subfigure{
\includegraphics[width=1.01in]{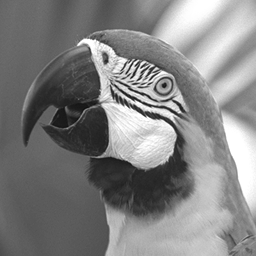}}
\hspace{0.06in}
\subfigure{
\includegraphics[width=1.01in]{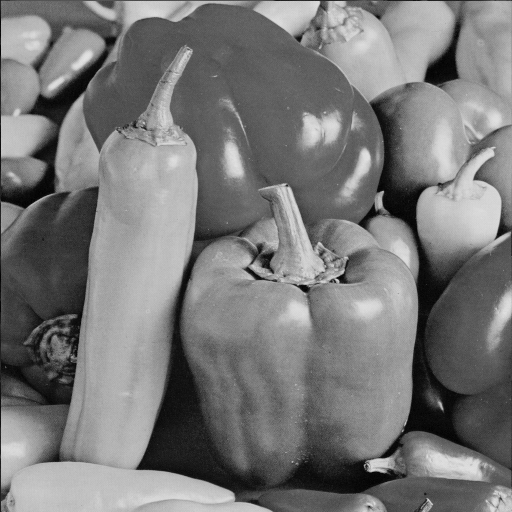}}
\hspace{0.06in}
\subfigure{
\includegraphics[width=1.01in]{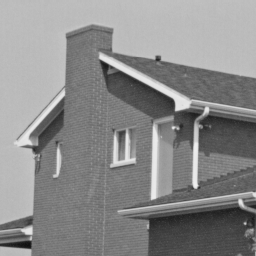}}
\hspace{0.06in}
\subfigure{
\includegraphics[width=1.01in]{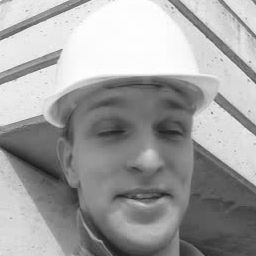}}
\label{Fig:test}
\vskip -0.065in \tiny{\quad Parrot \quad\quad\quad\quad\quad\quad\quad\quad\quad\quad Peppers \quad\quad\quad\quad\quad\quad\quad\quad\quad\quad House \quad\quad\quad\quad\quad\quad\quad\quad\quad\quad\quad Foreman}
\vskip -0.08in \caption{The thumbnails of eight test images.}
\label{Fig:a1}
\vspace{-0.28in}
\end{figure*}

For the local structural sampling process, each row of the sampling matrix $\Phi$ can be considered as a filter and therefore we can use a series of convolutional operations to simulate the sampling process~\cite{shi2019image}. Therefore, we propose a new local structural sampling network to sample the image, in which a convolutional layer is used to imitate the sampling process. Specifically, given the sampling ratio $\frac{M}{N}$, there are $n_{B}=\lfloor \frac{M}{N}B^{2}\rfloor$ convolutional filters in the sampling network, which correspond to the rows of the sampling matrix to obtain $n_{B}$ CS measurements. Since the size of each image block is $B\times B$, the size of each convolutional filter is also $B\times B$, and each filter outputs one measurement for each image block. Besides, it is worth noting that the stride of the convolutional layer is also $B\times B$ for non-overlapping sampling and there is no bias in each convolutional filter.

\begin{table*}[t]
\renewcommand\arraystretch{0.85}
\centering
\caption{The rate-distortion (PSNR) comparisons against other image CS coding algorithms in terms of three sampling ratios (R) 0.1, 0.2 and 0.3. Ours-H is the proposed CS coding method based on BPG (i.e., CSCNet-H). The bold text is the best performance.}
\vspace{-0.15in}
\label{Tab:a1}

\begin{tabular}{p{1.1cm}<{\centering} | p{0.38cm}<{\centering} || p{0.61cm}<{\centering} p{0.61cm}<{\centering} p{0.61cm}<{\centering} | p{0.61cm}<{\centering} p{0.61cm}<{\centering} p{0.61cm}<{\centering} | p{0.61cm}<{\centering} p{0.61cm}<{\centering} p{0.61cm}<{\centering} | p{0.61cm}<{\centering}  p{0.61cm}<{\centering} p{0.61cm}<{\centering}}
\hline
\multirow{2}*{\small{Images}} & \multirow{2}*{\small{Bpp}}  & \multicolumn{3}{c|}{\small{SDPC~\cite{Zhang2013Spatially}}} & \multicolumn{3}{c|}{\small{LSMM~\cite{gao2015block}}}  & \multicolumn{3}{c|}{\small{DQBCS~\cite{Cui2018An}}} & \multicolumn{3}{c}{\small{Ours-H}} \\
\cline{3-14}
&&\small{R=}0.1&\small{R=}0.2&\small{R=}0.3&\small{R=}0.1&\small{R=}0.2&\small{R=}0.3&\small{R=}0.1&\small{R=}0.2&\small{R=}0.3&\small{R=}0.1&\small{R=}0.2&\small{R=}0.3\\
\hline

\multirow{5}*{\small{Lenna}}&0.1&19.73&16.99&16.27&23.64&18.66&19.74 &24.91&24.73&25.65&\textbf{26.97} &\textbf{27.56}&\textbf{27.72}\\
&0.2&22.47&18.79&18.33&26.55&25.87&24.24 &27.43&27.39&27.20&\textbf{28.16} &\textbf{29.82}&\textbf{30.78}\\
&0.3&24.72&21.78&19.69&26.77&28.10&26.88 &28.05&28.58&28.69&\textbf{28.35} &\textbf{30.67}&\textbf{32.41}\\
&0.4&24.97&23.59&20.86&26.81&28.28&29.29 &28.10&29.49&29.67&\textbf{28.40} &\textbf{30.98}&\textbf{33.30}\\
&0.5&25.10&26.53&23.26&26.83&28.35&29.51 &28.13&29.86&30.46&\textbf{28.42} &\textbf{31.11}&\textbf{33.68}\\

\hline
\multirow{5}*{\small{Boats}}&0.1&19.97&16.54&15.90&23.66&21.30&18.88&23.67&23.48&23.56&\textbf{26.22} &\textbf{27.02}&\textbf{26.90}\\
&0.2&22.49&19.78&18.02&26.10&25.38&23.11&25.75&25.71&25.49&\textbf{27.73} &\textbf{29.48}&\textbf{29.88}\\
&0.3&24.07&22.20&19.49&26.67&27.48&27.05&26.10&26.65&26.76&\textbf{27.94} &\textbf{30.38}&\textbf{31.36}\\
&0.4&24.69&24.04&21.60&26.73&28.63&28.35&26.46&27.73&28.01&\textbf{28.01} &\textbf{30.77}&\textbf{32.18}\\
&0.5&24.79&26.30&23.64&26.75&28.76&29.21 &26.48&28.10&28.64&\textbf{28.02} &\textbf{30.92}&\textbf{32.65}\\

\hline
\multirow{5}*{\small{Barbara}}&0.1&19.17&18.16&17.20&22.06&21.16&19.03 &21.25&21.02&20.83&\textbf{22.29} &\textbf{23.74}&\textbf{24.11}\\
&0.2&22.17&19.64&19.03&\textbf{23.50}&23.24&20.99 &21.93&21.88&21.71&22.56 &\textbf{24.50}&\textbf{25.20}\\
&0.3&23.14&21.66&20.39&\textbf{23.73}&24.16&24.32 &22.31&22.10&22.21&22.60 &\textbf{24.82}&\textbf{25.75}\\
&0.4&23.39&23.18&21.22&\textbf{23.75}&24.33&24.94 &22.56&24.30&24.52&22.61 &\textbf{24.95}&\textbf{26.08}\\
&0.5&23.46&24.26&22.85&\textbf{23.77}&24.48&25.17 &22.60&24.77&25.20&22.61 &\textbf{25.03}&\textbf{26.25}\\

\hline
\multirow{5}*{\small{Monarch}}&0.1&16.92&15.77&15.34&20.59&18.06&17.50 &22.83&22.64&22.70&\textbf{24.66} &\textbf{24.78}&\textbf{24.91}\\
&0.2&19.90&17.46&17.19&23.49&22.52&19.50 &25.06&25.00&24.81&\textbf{26.18} &\textbf{27.88}&\textbf{28.44}\\
&0.3&21.08&20.23&18.58&23.95&25.41&23.50 &25.80&26.80&26.89&\textbf{26.65} &\textbf{29.21}&\textbf{30.40}\\
&0.4&21.40&21.59&19.62&24.03&26.30&25.80 &26.10&27.91&28.08&\textbf{26.72} &\textbf{29.81}&\textbf{31.52}\\
&0.5&21.59&23.93&21.28&24.05&26.83&27.11 &26.12&28.00&28.35&\textbf{26.73} &\textbf{30.07}&\textbf{32.16}\\

\hline
\multirow{5}*{\small{House}}&0.1&21.81&16.83&15.82&26.66&23.99&20.93 &28.43&28.25&28.25&\textbf{30.92} &\textbf{32.33}&\textbf{32.73}\\
&0.2&25.97&21.96&18.19&28.98&29.46&27.50 &30.34&30.32&30.09&\textbf{31.37} &\textbf{33.58}&\textbf{34.74}\\
&0.3&27.23&24.96&21.33&29.16&31.04&30.87 &31.12&31.81&31.90&\textbf{31.42} &\textbf{33.88}&\textbf{35.40}\\
&0.4&27.64&27.46&24.13&29.22&31.78&32.41 &31.31&32.39&32.56&\textbf{31.43} &\textbf{34.05}&\textbf{35.82}\\
&0.5&27.86&29.88&26.97&29.23&31.85&32.63 &31.34&32.50&32.93&\textbf{31.44} &\textbf{34.12}&\textbf{36.01}\\
\hline
\multirow{5}*{\small{Parrot}}&0.1&19.97&17.18&15.12&23.97&22.05&19.89 &25.17&24.98&25.04&\textbf{27.09} &\textbf{28.78}&\textbf{29.54}\\
&0.2&22.56&19.82&16.88&25.51&26.28&24.87 &26.70&26.66&26.45&\textbf{27.58} &\textbf{30.37}&\textbf{32.13}\\
&0.3&23.89&22.27&19.94&25.76&28.00&28.16 &27.18&27.69&27.82&\textbf{27.61} &\textbf{30.79}&\textbf{33.14}\\
&0.4&24.20&24.81&22.14&25.79&28.43&29.17 &27.30&29.33&29.53&\textbf{27.64} &\textbf{30.95}&\textbf{33.67}\\
&0.5&24.34&25.44&22.97&25.80&28.58&29.75 &27.33&29.51&29.71&\textbf{27.64} &\textbf{31.01}&\textbf{33.90}\\

\hline
\multirow{5}*{\small{Peppers}}&0.1&18.10&16.68&15.54&22.28&18.81&17.09 &24.15&23.92&24.00&\textbf{26.99} &\textbf{26.40}&\textbf{26.48}\\
&0.2&22.00&18.39&17.45&25.05&24.90&22.90 &26.00&25.94&25.77&\textbf{28.53} &\textbf{28.74}&\textbf{28.71}\\
&0.3&23.96&20.66&18.90&25.50&26.99&25.88 &27.20&27.68&27.81&\textbf{28.88} &\textbf{29.55}&\textbf{29.82}\\
&0.4&24.64&23.29&20.05&25.63&28.43&27.41 &27.31&28.84&29.00&\textbf{28.93} &\textbf{29.83}&\textbf{30.39}\\
&0.5&24.79&26.31&22.63&25.69&28.44&28.71 &27.34&29.16&29.56&\textbf{28.95} &\textbf{29.98}&\textbf{30.67}\\

\hline
\multirow{5}*{\small{Foreman}}&0.1&22.70&18.27&15.37&28.25&23.51&22.94 &27.66&27.45&27.53&\textbf{33.28} &\textbf{33.44}&\textbf{33.58}\\
&0.2&28.31&22.52&19.32&31.19&30.70&28.73 &29.01&31.63&31.45&\textbf{34.40} &\textbf{35.60}&\textbf{36.27}\\
&0.3&30.05&24.69&21.93&31.42&32.76&31.50 &31.59&32.90&32.08&\textbf{34.60} &\textbf{36.32}&\textbf{37.50}\\
&0.4&30.45&29.34&24.43&31.48&33.69&34.01 &31.92&33.41&33.49&\textbf{34.68} &\textbf{36.60}&\textbf{38.20}\\
&0.5&30.64&31.81&29.46&31.50&33.88&34.66 &31.94&33.79&34.12&\textbf{34.70} &\textbf{36.78}&\textbf{38.61}\\

\hline
\multirow{5}*{\small{Average}}&0.1&19.79&17.05&15.82&23.89&20.94&19.50 &24.76&24.56&24.67&\textbf{27.30} &\textbf{28.01}&\textbf{28.25}\\
&0.2&23.23&19.80&18.05&26.30&26.04&23.98 &26.53&26.82&26.62&\textbf{28.31} &\textbf{30.00}&\textbf{30.77}\\
&0.3&24.77&22.31&20.03&26.62&27.99&27.27 &27.42&28.03&28.15&\textbf{28.51} &\textbf{30.70}&\textbf{31.97}\\
&0.4&25.17&24.66&21.76&26.68&28.73&28.92 &27.68&29.18&29.36&\textbf{28.55} &\textbf{30.99}&\textbf{32.65}\\
&0.5&25.32&26.81&24.13&26.70&28.90&29.59 &27.73&29.46&29.87&\textbf{28.56} &\textbf{31.13}&\textbf{32.99}\\

\hline

\end{tabular}
\vspace{-0.12in}
\end{table*}

For the convenience of depiction, the convolutional filters of the sampling network are initialed as $W_{l}^{*}$ and the $k$-th filter is denoted by $W_{l}^{*}(k)$, where the index $k=\{1,2\cdots n_{B}\}$. In order to perform local structural sampling, a series of binary masks $\Upsilon(k)$ of the same size as $W_{l}^{*}(k)$ are first introduced, which can be represented by

\vspace{-0.15in}
\begin{equation}
\begin{split}
\Upsilon = \{ \Upsilon(1) \cdot \hspace{-0.03in} \cdot \hspace{-0.03in} \cdot   \Upsilon(n_{B})\}  \Leftarrow
\left\{
\begin{bmatrix}
I_{L\times L}  & \hskip -0.08in O_{L\times E}  \\
O_{E\times L}  & \hskip -0.08in O_{E\times E}  \\
\end{bmatrix}
\hskip -0.02in
\cdot \hspace{-0.03in} \cdot \hspace{-0.03in} \cdot
\hskip -0.02in
\begin{bmatrix}
O_{E\times E}  & \hskip -0.08in O_{E\times L}   \\
O_{L\times E}  & \hskip -0.08in I_{L\times L}  \\
\end{bmatrix}
 \right\}
\end{split}
\label{Eq:a4}
\end{equation}
where $\Upsilon(k)$ indicates the $k$-th mask of $\Upsilon$. $I$ indicates the all-1 submatrix and $O$ is the all-0 submatrix. The subscript represents the dimension of the submatrix and $E=B-L$. In fact, the all-1 submatrices can be regarded as the sliding windows ($L\times L$), which slide on the corresponding masks with the increase of the index $k$. By introducing binary mask $\Upsilon$, the localization operation upon the convolutional filters can be performed as

\vspace{-0.1in}
\begin{equation}
\tilde{W}_{l}=\Upsilon\circ W_{l}^{*}
\label{Eq:a5}
\end{equation}
where $\circ$ indicates the element-wise multiplication. For the $k$-th filter, the localization operation can be expressed as $\tilde{W}_{l}(k)=\Upsilon(k)\circ W_{l}^{*}(k)$, which in fact signifies that only $L^{2}$ values of $W_{l}^{*}(k)$ corresponding to the all-1 submatrix of $\Upsilon(k)$ are preserved in $\tilde{W}_{l}(k)$, and the others are set to zero. Apparently, through the above localization operation, most elements of the proposed sampling matrix are zero, which signifies that the designed local sampling matrix is highly sparse.

Furthermore, to limit the range of the produced measurements, an additional positive normalization operation is introduced, in which a positive mapping function ($F_{p}$) and a normalization operator ($S_{n}$) are employed. The positive normalization operation can be expressed as

\vspace{-0.08in}
\begin{equation}
W_{l}(k) = S_{n}(F_{p}(\tilde{W}_{l}(k)))
\label{Eq:a6}
\end{equation}
where $F_{p}$ is responsible for converting the retained $L^{2}$ elements in $\tilde{W}_{l}(k)$ into the positive ones and $S_{n}$ further constrains the sum of these $L^{2}$ positive elements is approximately equal to 1. It is noted that through the positive normalization operation, the range of the generated measurements is identical with the intensities of natural images.

After the above operations, the normalized convolutional filters $W_{l}$=$\{W_{l}(1), W_{l}(2), \cdots, W_{l}(n_{B})\}$ are generated, and the proposed local structural sampling process can be expressed as:
\begin{equation}
y = W_{l} \ast x
\label{Eq:1}
\end{equation}
where $\ast$ represents convolutional operation. $W_{l}$ indicate the learnable weights of filters (actually is the sampling matrix $\Phi$) in the sampling network, which are jointly optimized with the other functional modules. The output $y$ signifies the produced CS measurements and for each column of $y$, there are $n_{B}$ measurements in terms of one image block.

Fig.~\ref{Fig:local_sampling} shows more detailed intuitive explanations of our proposed local structural sampling process. Apparently, the introduced localization operation ensures local non-zero characteristic of the sampling matrix, and the positive normalization operation constrains the range of generated measurements. Clearly, by introducing these two operations, each row of our proposed local structural sampling matrix only perceives the local position of the given image blocks. Based on above local perception mode and the local smoothing property of natural images, a high correlation between the generated measurements is preserved in most cases. Besides, the proposed local structural sampling matrix can be jointly trained with other functional modules in an end-to-end fashion, and it is worth emphasizing that this optimization strategy directly learns the optimal sampling matrix from a large amount of training data, rather than constructing sampling matrix based on RIP property. Therefore, this sampling matrix learning method usually does not need to pay attention to RIP property, and the end-to-end training mode mostly ensures the effectiveness of the learned sampling matrix.

\begin{table*}[t]
\renewcommand\arraystretch{0.88}
\centering
\caption{The rate-distortion (SSIM) comparisons against other image CS coding algorithms in terms of three sampling ratios (R) 0.1, 0.2 and 0.3. Ours-H is the proposed CS coding method based on BPG (i.e., CSCNet-H). The bold text is the best performance.}
\vspace{-0.15in}
\label{Tab:a2}

\begin{tabular}{p{0.6cm}<{\centering} || p{0.72cm}<{\centering} p{0.72cm}<{\centering} p{0.72cm}<{\centering} | p{0.72cm}<{\centering} p{0.72cm}<{\centering} p{0.72cm}<{\centering} | p{0.72cm}<{\centering} p{0.72cm}<{\centering} p{0.72cm}<{\centering} | p{0.72cm}<{\centering}  p{0.72cm}<{\centering} p{0.72cm}<{\centering}}
\hline
\multirow{2}*{Bpp}  & \multicolumn{3}{c|}{\small{SDPC~\cite{Zhang2013Spatially}}} & \multicolumn{3}{c|}{\small{LSMM~\cite{gao2015block}}}  & \multicolumn{3}{c|}{\small{DQBCS~\cite{Cui2018An}}} & \multicolumn{3}{c}{\small{Ours-H}} \\
\cline{2-13}
\vspace{0.06in}
&\small{R=}0.1&\small{R=}0.2&\small{R=}0.3&\small{R=}0.1&\small{R=}0.2&\small{R=}0.3&\small{R=}0.1&\small{R=}0.2&\small{R=}0.3&\small{R=}0.1&\small{R=}0.2&\small{R=}0.3\\
\hline
0.1&0.344&0.172&0.114&0.696&0.600&0.527 &0.716&0.687&0.686&\textbf{0.801} &\textbf{0.804}&\textbf{0.809}\\
0.2&0.593&0.312&0.201&0.787&0.761&0.694 &0.801&0.796&0.788&\textbf{0.851} &\textbf{0.869}&\textbf{0.878}\\
0.3&0.713&0.462&0.296&0.803&0.822&0.789 &0.814&0.830&0.838&\textbf{0.864} &\textbf{0.897}&\textbf{0.909}\\
0.4&0.742&0.598&0.400&0.805&0.846&0.836 &0.828&0.853&0.865&\textbf{0.868} &\textbf{0.909}&\textbf{0.926}\\
0.5&0.751&0.706&0.503&0.807&0.855&0.862 &0.830&0.871&0.884&\textbf{0.869} &\textbf{0.916}&\textbf{0.935}\\
\hline

\end{tabular}
\vspace{-0.14in}
\end{table*}

\subsection{Measurement Coding}
\label{section:a32}

After sampling process, a measurement coding module is developed to compress the produced measurements into bitstreams. According to the local smoothness property of the natural images, the high correlations between the adjacent measurements are preserved after sampling process~\cite{gao2015block}. According to the spatial position of the local nonzero windows in $W_{l}$, we first transform the 1D measurements $y_{(i,j)}$ into 2D tensors for each image block $x_{(i,j)}$ and then concatenate them together (purple block of Fig.~\ref{Fig:reconstruction}). The transformation and concatenation processes can be expressed as:
\begin{equation}
\begin{split}
\hat{y}=\hspace{-0.03in}{\rm Cat}
\hspace{-0.01in}\begin{pmatrix}
R_{t}(y_{(1,1)})  & \hspace{-0.1in} R_{t}(y_{(1,2)}) & \hspace{-0.1in} \cdots  & \hspace{-0.1in} R_{t}(y_{(1,\frac{h}{B})})   \\
R_{t}(y_{(2,1)})  & \hspace{-0.1in} R_{t}(y_{(2,2)}) & \hspace{-0.1in} \cdots  & \hspace{-0.1in} R_{t}(y_{(2,\frac{h}{B})})  \\
\vdots  & \vdots & \ddots  & \vdots  \\
R_{t}(y_{(\frac{w}{B},1)})  & \hspace{-0.08in} R_{t}(y_{(\frac{w}{B},2)}) & \hspace{-0.1in} \cdots  & \hspace{-0.1in} R_{t}(y_{(\frac{w}{B},\frac{h}{B})})  \\
\end{pmatrix}
\end{split}
\label{Eq:28}
\end{equation}
where $R_{t}(\cdot)$ is the transform operator and $\rm Cat(\cdot)$ is the concatenation function, after which the integrated measurements $\hat{y}$ of the whole image $x$ are produced. For more details of the operator $R_{t}(\cdot)$, we try to convert the measurement $y_{(i,j)}$ into a square shape. But the number of measurement (i.e., $n_{B}$) usually cannot be expressed as the square of an integer. To solve above problem, we pad the blank measurement according to the adjacent measurement values in our experiments.

\begin{table*}[t]
\renewcommand\arraystretch{0.85}
\centering
\caption{The rate-distortion comparisons (PSNR and SSIM) with other CS-based image coding algorithms. Ours-J-0.25, Ours-H-0.25 and Ours-H-0.50 are the proposed CS coding methods (i.e., CSCNet-J-0.25, CSCNet-H-0.25, CSCNet-H-0.50) based on JPEG2000 and BPG, and the sampling ratios are 0.25 and 0.50.}
\vspace{-0.15in}
\label{Tab:a3}

\begin{tabular}{p{1.10cm}<{\centering} | p{0.38cm}<{\centering} || p{0.56cm}<{\centering}  p{0.59cm}<{\centering} | p{0.56cm}<{\centering}  p{0.59cm}<{\centering} | p{0.56cm}<{\centering}  p{0.59cm}<{\centering} || p{0.56cm}<{\centering}  p{0.59cm}<{\centering} | p{0.56cm}<{\centering}  p{0.59cm}<{\centering} | p{0.56cm}<{\centering}  p{0.59cm}<{\centering}}
\hline

\multirow{2}*{\small{Images}} & \multirow{2}*{\small{Bpp}} & \multicolumn{2}{c|}{\small{LRS-J-0.25}} & \multicolumn{2}{c|}{\small{LRS-H-0.25}} & \multicolumn{2}{c||}{\small{DLAMP-CS}} & \multicolumn{2}{c|}{\small{Ours-J-0.25}} & \multicolumn{2}{c|}{\small{Ours-H-0.25}} & \multicolumn{2}{c}{\small{Ours-H-0.50}} \\
\cline{3-14}
&&\small{PSNR}&\small{SSIM}&\small{PSNR}&\small{SSIM}&\small{PSNR}&\small{SSIM}&\small{PSNR}&\small{SSIM}&\small{PSNR}&\small{SSIM}&\small{PSNR}&\small{SSIM}\\
\hline
\multirow{5}*{\small{Lenna}}&0.1&25.57&0.743&26.91&0.785&26.09&0.766&25.82&0.745 &\textbf{27.30}&\textbf{0.801}&27.20&0.794\\
&0.2&28.23&0.819&29.18&0.849&29.85&0.860&28.82&0.842 &30.45&0.880&\textbf{30.70}&\textbf{0.881}\\
&0.3&30.06&0.867&30.13&0.871&32.04&0.898&30.32&0.885 &32.10&0.920&\textbf{33.02}&\textbf{0.922}\\
&0.4&30.58&0.881&31.23&0.905&33.95&0.922&31.15&0.910 &32.80&0.936&\textbf{34.41}&\textbf{0.940}\\
&0.5&31.08&0.896&31.68&0.921&35.12&0.934&31.52&0.920 &33.09&0.944&\textbf{35.25}&\textbf{0.951}\\

\hline
\multirow{5}*{\small{Boats}}&0.1&25.36&0.701&26.33&0.737&24.79&0.666&25.38&0.713&27.08&\textbf{0.773}&\textbf{27.11}&0.772\\
&0.2&28.03&0.792&28.93&0.825&29.10&0.823&28.30&0.807&\textbf{29.91}&\textbf{0.862}&29.76&0.853\\
&0.3&29.55&0.841&29.74&0.850&31.10&0.877&29.86&0.863&31.49&\textbf{0.903}&\textbf{31.74}&\textbf{0.903}\\
&0.4&30.41&0.862&30.90&0.885&32.96&0.909&30.81&0.891&32.37&\textbf{0.924}&\textbf{33.01}&0.921\\
&0.5&30.92&0.882&31.45&0.903&33.94&0.921&31.26&0.905 &32.75&0.934&34.02&\textbf{0.937}\\

\hline
\multirow{5}*{\small{Barbara}}&0.1&23.26&0.627&23.89&0.655&22.81&0.565&23.57&0.647 &24.00&0.703&\textbf{25.99}&\textbf{0.758}\\
&0.2&24.23&0.696&24.67&0.714&25.92&0.718&24.89&0.759 &24.98&0.795&\textbf{28.15}&\textbf{0.847}\\
&0.3&25.03&0.765&25.41&0.759&27.87&0.835&25.55&0.810 &25.33&0.825&29.98&\textbf{0.896}\\
&0.4&25.42&0.786&25.97&0.810&29.72&0.882&25.81&0.826 &25.50&0.845&31.43&0.925\\
&0.5&25.62&0.807&26.12&0.827&31.42&0.911&26.02&0.841 &25.67&0.855&32.26&0.939\\

\hline
\multirow{5}*{\small{Monarch}}&0.1&22.13&0.698&24.50&0.791&22.43&0.679&22.89&0.733 &\textbf{25.07}&\textbf{0.805}&24.87&0.800\\
&0.2&25.51&0.817&27.26&0.868&27.45&0.848&26.20&0.836 &\textbf{28.14}&\textbf{0.885}&27.92&0.880\\
&0.3&27.66&0.867&28.82&0.900&30.25&0.911&28.46&0.895 &30.23&0.922&\textbf{30.35}&\textbf{0.923}\\
&0.4&29.02&0.901&29.81&0.924&31.92&0.934&29.32&0.915 &31.56&\textbf{0.943}&\textbf{32.15}&\textbf{0.943}\\
&0.5&29.67&0.912&30.62&0.941&\textbf{33.65}&0.947&29.92&0.931 &32.20&0.954&33.10&\textbf{0.955}\\

\hline
\multirow{5}*{\small{House}}&0.1&29.74&0.798&31.75&0.842&31.05&0.824&29.55&0.809 &\textbf{32.96}&\textbf{0.855}&32.70&0.851\\
&0.2&32.19&0.833&32.95&0.856&34.48&0.869&32.79&0.855 &34.70&\textbf{0.884}&\textbf{34.92}&0.880\\
&0.3&33.30&0.852&33.32&0.861&35.79&0.882&33.72&0.878 &35.41&0.900&\textbf{35.92}&\textbf{0.903}\\
&0.4&33.66&0.858&34.25&0.880&36.44&0.890&33.98&0.889 &35.77&0.910&\textbf{36.70}&\textbf{0.917}\\
&0.5&33.91&0.870&34.50&0.890&37.24&0.896&34.07&0.895 &36.01&0.914&37.11&\textbf{0.927}\\
\hline
\multirow{5}*{\small{Parrot}}&0.1&26.90&0.809&28.15&0.844&28.17&0.824&27.32&0.817 &\textbf{29.40}&\textbf{0.857}&29.33&0.855\\
&0.2&29.17&0.856&29.90&0.879&32.12&0.891&30.24&0.886 &31.92&\textbf{0.912}&\textbf{32.80}&0.910\\
&0.3&30.19&0.886&30.29&0.889&34.07&0.910&31.32&0.921 &32.80&\textbf{0.935}&\textbf{34.48}&0.932\\
&0.4&30.48&0.896&30.86&0.917&\textbf{35.58}&0.927&31.52&0.929 &33.33&0.946&35.55&\textbf{0.948}\\
&0.5&30.61&0.908&30.98&0.928&\textbf{36.46}&0.933&31.67&0.936 &33.43&0.951&36.35&\textbf{0.957}\\

\hline
\multirow{5}*{\small{Peppers}}&0.1&23.90&0.726&24.97&0.785&24.17&0.710&24.78&0.743 &\textbf{26.20}&\textbf{0.792}&26.17&0.782\\
&0.2&26.76&0.825&27.02&0.852&\textbf{29.26}&0.845&27.37&0.836 &28.91&\textbf{0.875}&28.78&0.866\\
&0.3&27.90&0.861&27.85&0.873&\textbf{31.73}&0.888&28.60&0.876 &29.93&\textbf{0.901}&30.18&\textbf{0.901}\\
&0.4&28.69&0.879&28.65&0.901&\textbf{33.35}&0.909&29.19&0.894 &30.50&\textbf{0.920}&30.89&0.914\\
&0.5&29.06&0.892&28.94&0.914&\textbf{34.38}&0.919&29.41&0.905 &30.72&\textbf{0.929}&31.36&0.927\\

\hline
\multirow{5}*{\small{Foreman}}&0.1&30.78&0.847&32.79&0.888&32.18&0.872&31.45&0.864 &\textbf{33.81}&\textbf{0.897}&33.76&0.895\\
&0.2&33.76&0.885&34.67&0.910&35.89&0.921&34.11&0.905 &\textbf{36.39}&\textbf{0.932}&36.19&0.928\\
&0.3&35.86&0.909&35.14&0.916&37.34&0.934&35.72&0.931 &37.55&\textbf{0.948}&\textbf{37.70}&0.945\\
&0.4&36.26&0.915&36.86&0.936&38.19&0.941&36.12&0.939 &38.13&0.956&\textbf{38.76}&\textbf{0.959}\\
&0.5&36.70&0.926&37.48&0.946&38.91&0.953&36.48&0.947 &38.44&0.959&\textbf{39.13}&\textbf{0.962}\\

\hline
\multirow{5}*{\small{Average}}&0.1&25.96&0.744&27.41&0.791&26.46&0.736&26.35&0.759 &28.23&0.811&\textbf{28.39}&\textbf{0.813}\\
&0.2&28.49&0.815&29.32&0.844&30.51&0.846&29.09&0.841 &30.68&0.878&\textbf{31.15}&\textbf{0.881}\\
&0.3&29.94&0.856&30.09&0.865&32.52&0.892&30.44&0.882 &31.86&0.907&\textbf{32.92}&\textbf{0.916}\\
&0.4&30.57&0.872&31.07&0.895&34.02&0.914&30.99&0.899 &32.50&0.923&\textbf{34.11}&\textbf{0.934}\\
&0.5&30.95&0.887&31.47&0.909&\textbf{35.14}&0.927&31.29&0.910 &32.79&0.930&34.83&\textbf{0.944}\\

\hline

\end{tabular}
\vspace{-0.15in}
\end{table*}

Through the above transformation and concatenation operations, a high correlation between the spatially adjacent measurements in $\hat{y}$ is preserved. Because the range of the measurements is limited identical with the intensities of the images, it is reasonable to encode these measurements by an existing image codec. In our framework, a third-party image coding scheme is utilized to compress the produced measurements. The measurement coding process can be denoted by
\begin{equation}
\tilde{y} = {\rm Dec}({\rm Enc}(\hat{y}))
\label{hahaha}
\end{equation}
where ${\rm Enc(\cdot)}$ and ${\rm Dec(\cdot)}$ indicate encoder and decoder processes of the third-party image coding scheme respectively. $\tilde{y}$ is the produced decoded measurements. Using the existing third-party image codec to compress measurements not only saves the design cost of the measurement coding tools, but also enhances the flexibility of the proposed model.

\subsection{Laplacian Pyramid Reconstruction Network}
\label{section:a33}

The existing deep network-based CS reconstruction algorithms~\cite{shi2019image,shi2019scalable} usually reconstruct the target image in a single scale space, and all convolutional operations are carried out on the scale space of the same size as the original image, which not only limits the CS reconstruction quality, but also increases the computational complexity to a certain extent. To relieve this problem, in the reconstruction module, a Laplacian pyramid reconstruction network is proposed as shown in Fig.~\ref{Fig:reconstruction} to reconstruct the target image from the measurement domain to the image domain. Specifically, in the proposed reconstruction network, an initial reconstruction with lower resolution is first obtained through an upsampling operator. Then, a convolutional Laplacian pyramid architecture with multiple levels is followed to extract and aggregate multi-scale feature representations. After that, a series of residual blocks are appended for feature enhancement. At last, a convolutional layer is used to reveal the final reconstructed image.

Specifically, for the initial reconstruction process, we first execute a bilinear interpolation upsampling upon the decoded measurements $\tilde{y}$ to produce the initial reconstruction $F_{0}$ of size $\frac{w}{S}\times \frac{h}{S}$, where $S$ is a well-designed scale factor. Apparently, it is necessary to minimize information loss as much as possible in the initial reconstruction to ensure the reconstructed quality of the target image. Therefore, we need to guarantee that the dimension of the initial reconstruction keeps higher than that of the decoded measurements. On the other hand, the initial reconstruction needs to be small enough to ensure a larger number of pyramid levels in the proposed pyramid architecture, realizing efficient extraction and aggregation of multi-scale deep representations. As mentioned above, the scale factor $S$ can be calculated by the following steps. Let $\mathcal{Q} = \{2^i | i=0, 1, 2, \ldots \}$ and $\mathcal{P} = \{j| j\in \mathcal{Q} \quad and \quad \frac{1}{2^j} > \frac{M}{N}\}$. The optimal scale factor $S$ can be obtained by
\vspace{-0.01in}
\begin{eqnarray}
\label{qeu2}
  S = {\rm Max}(\mathcal{P})
\vspace{-0.01in}
\end{eqnarray}
where the function ${\rm Max}(\cdot)$ is used to return the maximal element of a given set.

After the initial reconstruction, a convolutional Laplacian pyramid architecture with $log_{2}S$ levels is further performed to aggregate multi-scale features and the network structure is shown in Fig.~\ref{Fig:reconstruction}. For each level, two submodules are included, i.e., feature extraction and feature upsampling. Specifically, for the feature extraction submodule of each level, several residual blocks are included and the network architecture of the residual block is shown in Fig.~\ref{Fig:reconstruction} (gray region). After the feature extraction submodule, a feature upsampling submodule is appended for the transformation between the different scale spaces, in which a deconvolutional layer (DConv) is involved and the size of its filter is $4\times 4$. After all the Laplacian pyramid levels, a series of ($d$) residual blocks are stacked (gray blocks of Fig.~\ref{Fig:reconstruction}). Finally, a convolutional layer with single kernel is followed to reveal the final reconstruction. For the configuration details of the residual blocks in the reconstruction network, each convolutional layer consists of $q$ kernels with size of $3\times 3\times q$. For simplicity, we use $F_{c}(\cdot)$ to signify the operations of the reconstruction network, and use $\theta$ to represent its learnable parameters.

\section{EXPERIMENTAL RESULTS}
\label{section:a4}

In this section, we first demonstrate the loss function, and then elaborate the experimental results and implementation details as well as the extensive comparisons against the existing state-of-the-art methods.

\subsection{Loss Function}

In the field of image compression, two optimization factors are usually concerned. One is the reconstruction quality and the other is the rate constraint for less storage requirement. Therefore, the overall loss function of the proposed CS coding framework can be denoted by

\vskip -0.14in
\begin{equation}
\mathcal{L} = \mathcal{L}_{c} + \gamma\mathcal{L}_{r}
\end{equation}
where $\mathcal{L}_{c}$ and $\mathcal{L}_{r}$ are the reconstruction loss and rate constraint loss respectively. $\gamma$ is a hyper parameter to control the rate constraint loss item. Specifically, for the reconstruction loss item, the mission is to minimize the gap between the reconstructed output image and the input image, i.e.,
\vspace{-0.01in}
\begin{equation}
\mathcal{L}_{c} = \hspace{-0.03in} \frac{1}{2K}\hspace{-0.03in} \sum_{i=1}^{K}\| F_{c}(\hspace{-0.01in} \tilde{y}_{i}, \theta)\hspace{-0.02in} -\hspace{-0.02in} x_{i}\|_{F}^{2}
\end{equation}
where $\tilde{y}_{i}$ indicate the decoded measurements of image $x_{i}$ as shown in Eq.~\ref{hahaha}. $F_{c}(\cdot)$ indicates the proposed reconstruction network and $\theta$ signify its learnable parameters.

For the rate constraint loss item, we directly restrict the bit rate of the measurements in continuous measurement domain, rather than the bitstream domain. In fact, the correlations between the spatial adjacent measurements directly determine the coding efficiency of the third-party image codec. We employ a regularizer based on the total variation (TV) to represent the spatial correlations among the produced measurements for rate constraint. Specifically, given the generated measurements $\hat{y}$, the rate constraint loss item can be expressed as

\vspace{-0.06in}
\begin{equation}
\label{rate}
\mathcal{L}_{r} = \sum_{i,j}((\hat{y}_{i,j+1}-\hat{y}_{i,j})^{2}+(\hat{y}_{i+1,j}-\hat{y}_{i,j})^{2})^{\frac{\beta}{2}}
\vspace{-0.02in}
\end{equation}
where $i, j$ are the indexes of the measurements $\hat{y}$. $\beta$ is a hyper parameter in the TV regularizer.

In the training process, the normalized local structural sampling matrix (namely $W_{l}$) is first used to sample (convolutional operations) the input image in the forward phase and then we can obtain the gradient tensor $d_{W_{l}}$ of $W_{l}$ in the backward phase. Considering the optimization of local structural sampling matrix, we do not update $W_{l}$ directly, but update $W_{l}^{*}$ according to the gradient back propagation. In other words, we optimize the convolutional filter $W_{l}$ (namely  $\Phi$) by optimizing $W_{l}^{*}$ and we get the final learned local structural sampling matrix through $\Phi =W_{l}={\mathcal{F}}(\Upsilon \circ W_{l}^{*})$, where $\mathcal{F}(\cdot)$ indicates the mixture of functions $F_{p}$ and $S_{n}$. 

\begin{figure*}[t]
\hspace{-0.13in}
\begin{minipage}[t]{0.17\textwidth}
\centering
\includegraphics[width=0.99in]{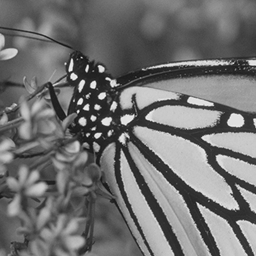}
\begin{tiny}
\centering

\end{tiny}
\end{minipage}
\hskip 0.3 cm
\begin{minipage}[t]{0.17\textwidth}
\centering
\includegraphics[width=0.99in]{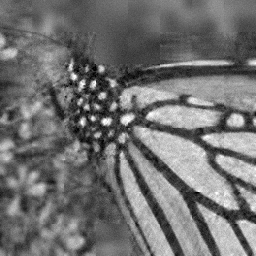}
\begin{scriptsize}
\centering
\end{scriptsize}
\end{minipage}
\hskip 0.3 cm
\begin{minipage}[t]{0.17\textwidth}
\centering
\includegraphics[width=0.99in]{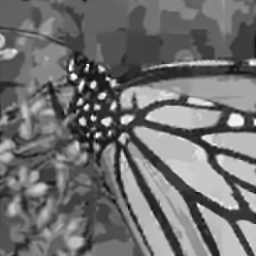}
\begin{scriptsize}
\centering
\end{scriptsize}
\end{minipage}
\hskip 0.3 cm
\begin{minipage}[t]{0.17\textwidth}
\centering
\includegraphics[width=0.99in]{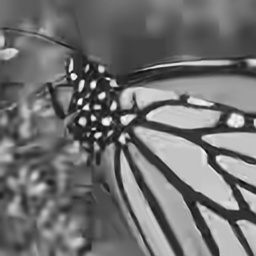}
\begin{scriptsize}
\centering
\end{scriptsize}
\end{minipage}
\hskip 0.3 cm
\begin{minipage}[t]{0.17\textwidth}
\centering
\includegraphics[width=0.99in]{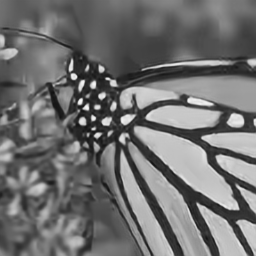}
\begin{scriptsize}
\centering
\end{scriptsize}
\end{minipage}
\\
\vskip -0.4 cm \hskip 0 cm \begin{tiny}Ground Truth \quad \quad \quad \quad \quad \quad \quad \quad \ SDPC \quad \quad \quad \quad \quad \quad \quad \quad \quad \quad \  LSMM \quad \quad \quad \quad \quad \quad \quad \quad \quad \quad DQBCS \quad \quad \quad \quad \quad \quad \quad \quad \ \quad CSCNet-H\end{tiny}
\vskip -0.13in
\caption{Visual quality comparisons between the proposed method and other image CS coding schemes in terms of R=0.2 and Bpp=0.2 on image \emph{Monarch}.}
\label{Fig:a111}
\vskip -0.03in
\end{figure*}

\begin{figure*}[t]
\hspace{-0.13in}
\begin{minipage}[t]{0.17\textwidth}
\centering
\includegraphics[width=0.99in]{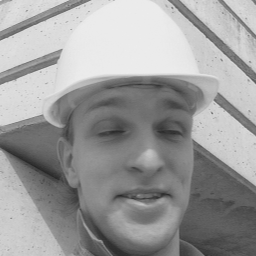}
\begin{tiny}
\centering

\end{tiny}
\end{minipage}
\hskip 0.3 cm
\begin{minipage}[t]{0.17\textwidth}
\centering
\includegraphics[width=0.99in]{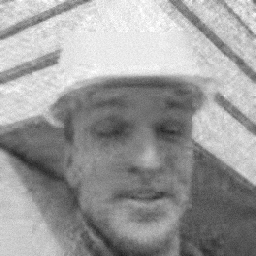}
\begin{scriptsize}
\centering
\end{scriptsize}
\end{minipage}
\hskip 0.3 cm
\begin{minipage}[t]{0.17\textwidth}
\centering
\includegraphics[width=0.99in]{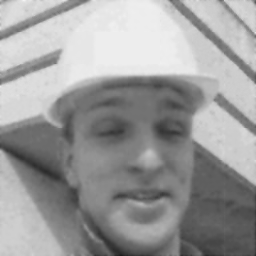}
\begin{scriptsize}
\centering
\end{scriptsize}
\end{minipage}
\hskip 0.3 cm
\begin{minipage}[t]{0.17\textwidth}
\centering
\includegraphics[width=0.99in]{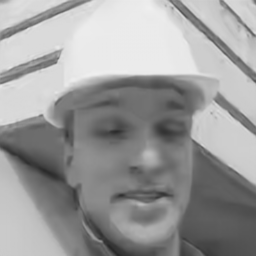}
\begin{scriptsize}
\centering
\end{scriptsize}
\end{minipage}
\hskip 0.3 cm
\begin{minipage}[t]{0.17\textwidth}
\centering
\includegraphics[width=0.99in]{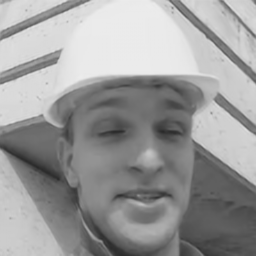}
\begin{scriptsize}
\centering
\end{scriptsize}
\end{minipage}
\\
\vskip -0.4 cm \hskip 0 cm \begin{tiny}Ground Truth \quad \quad \quad \quad \quad \quad \quad \quad \ SDPC \quad \quad \quad \quad \quad \quad \quad \quad \quad \quad \  LSMM \quad \quad \quad \quad \quad \quad \quad \quad \quad \quad DQBCS \quad \quad \quad \quad \quad \quad \quad \quad \ \quad CSCNet-H\end{tiny}

\vskip -0.07in
\vspace{-0.05in} \caption{Visual quality comparisons between the proposed method and other image CS coding schemes in terms of R=0.2 and Bpp=0.3 on image \emph{Foreman}.}
\label{Fig:a222}
\vskip -0.1in
\end{figure*}

\subsection{Implementation and Training Details}
\label{haha}

In the training process of the proposed CS coding framework CSCNet, we set block size $B$=32 as in~\cite{Cui2018An,shi2019scalable}. For the loss function, we set $\gamma$=0.1 and $\beta$=2. For functions $F_{p}$ and $S_{n}$, we first take the square of the nonzero elements of $\tilde{W}_{l}(k)$ for positive mapping ($F_{p}$), and then divide each positive element by their sum for summation normalization ($S_{n}$). In the proposed Laplacian pyramid reconstruction network (shown in Fig.~\ref{Fig:reconstruction}), we set $q=64$, $d=6$, and for the number of residual blocks in each level, we observe that with the increase of the residual block number, the reconstruction performance gradually converges and becomes stable. To balance the reconstruction performance and the computational complexity, we set the number of residual blocks in each level as 5 in our experiments, and the structure of the residual block is shown in Fig.~\ref{Fig:reconstruction}. For the local window size $L$ in sampling module, we find that when the size of local window is too large, the sampling matrix tends to perceive more larger area of the given image blocks, which reduces the correlation between the generated measurements, thereby weakening the measurement coding efficiency. On the other hand, when the size of local window is too small, the number of learnable parameters in the proposed local sampling matrix decreases, which limits the optimizability of the sampling matrix, thus affecting the sampling efficiency. As above, we set local window size $L$=3 as in~\cite{gao2015block}. We initialize the convolutional filters using the same method as~\cite{he2015delving} and pad zeros around the boundaries to keep the size of all feature maps the same as the input.

We use the training images of BSDS500~\cite{shi2019image} dataset and the training set of VOC2012~\cite{everingham2011pascal} as our training data. Specifically, in the training process, we use a batch size of 8 and randomly crop the size of patches to 128$\times$128. We augment the training data in the following two ways: ($1$) Rotate the images by 90$^{\circ}$, 180$^{\circ}$ and 270$^{\circ}$ randomly. ($2$) Flip the images horizontally with a probability of 0.5. We use the PyTorch deep learning toolbox and train our model using the Adaptive moment estimation (Adam) solver on a NVIDIA GTX 1080Ti GPU. We set the momentum to 0.9 and the weight decay to 1e-4. The learning rate is initialized to 1e-4 for all layers and decreased by a factor of 2 for every 30 epochs. We train our model for 200 epochs totally and 1000 iterations are performed for each epoch. Therefore 200$\times$1000 iterations are completed for the whole training process.

\begin{table*}[t]
\renewcommand\arraystretch{0.92}
\centering
\caption{The rate-distortion comparisons between the proposed CSCNet and the corresponding third-party image codecs.}
\vspace{-0.15in}
\label{Tab:ccc}
\begin{tabular}{p{0.5cm}<{\centering} || p{0.71cm}<{\centering}  p{0.71cm}<{\centering} | p{0.71cm}<{\centering}  p{0.71cm}<{\centering} | p{0.71cm}<{\centering}  p{0.71cm}<{\centering} || p{0.71cm}<{\centering}  p{0.71cm}<{\centering} | p{0.71cm}<{\centering}  p{0.71cm}<{\centering} | p{0.71cm}<{\centering}  p{0.71cm}<{\centering}}
\toprule
\multirow{3}*{\small{Bpp}} & \multicolumn{2}{c|}{\multirow{2}{*}{\small{JPEG2000}}} & \multicolumn{4}{c||}{\small{JPEG2000-based CSCNet}} & \multicolumn{2}{c|}{\multirow{2}{*}{\small{BPG}}} & \multicolumn{4}{c}{\small{BPG-based CSCNet}}\\
&&&\multicolumn{2}{c}{R=0.25} & \multicolumn{2}{c||}{R=0.50}
&&&\multicolumn{2}{c}{R=0.25} & \multicolumn{2}{c}{R=0.50} \\
\cline{2-13}
&\small{PSNR}&\small{SSIM}&\small{PSNR}&\small{SSIM}&\small{PSNR}&\small{SSIM}&\small{PSNR}&\small{SSIM}&\small{PSNR}&\small{SSIM}&\small{PSNR}&\small{SSIM}\\
\midrule
0.1&25.03&0.711&\textbf{26.35}&\textbf{0.759}&26.29&0.757 &27.88&0.793&28.23&0.811&\textbf{28.39}&\textbf{0.813}\\
0.2&28.53&0.813&29.09&0.841&\textbf{29.47}&\textbf{0.843} &\textbf{31.28}&0.876&30.68&0.878&31.15&\textbf{0.881}\\
0.3&30.73&0.863&30.44&0.882&\textbf{31.39}&\textbf{0.884} &\textbf{33.50}&0.913&31.86&0.907&32.92&\textbf{0.916}\\
0.4&32.48&0.894 &30.99&0.899&\textbf{32.56}&\textbf{0.910} &\textbf{35.08}&0.933 &32.50&0.923&34.11&\textbf{0.934}\\
0.5&\textbf{33.79}&0.912&31.29&0.910 &33.48&\textbf{0.925} &\textbf{36.40}&\textbf{0.945}&32.79&0.930 &34.83&0.944 \\

\bottomrule

\end{tabular}
\vspace{-0.03in}
\end{table*}

\begin{figure*}[t]
\vspace{-0.03in}
\hspace{-0.09in}
\begin{minipage}[t]{0.22\textwidth}
\centering
\includegraphics[width=1.26in]{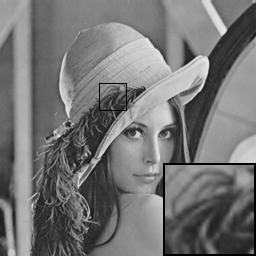}
\begin{tiny}
\centering
\end{tiny}
\end{minipage}
\hskip 0.3 cm
\begin{minipage}[t]{0.22\textwidth}
\centering
\includegraphics[width=1.26in]{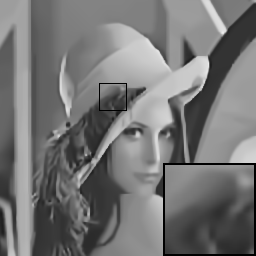}
\begin{scriptsize}
\centering
\end{scriptsize}
\end{minipage}
\hskip 0.3 cm
\begin{minipage}[t]{0.22\textwidth}
\centering
\includegraphics[width=1.26in]{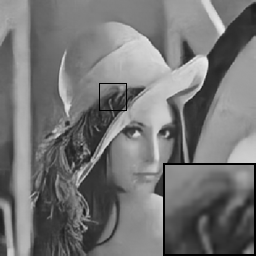}
\begin{scriptsize}
\centering
\end{scriptsize}
\end{minipage}
\hskip 0.3 cm
\begin{minipage}[t]{0.22\textwidth}
\centering
\includegraphics[width=1.26in]{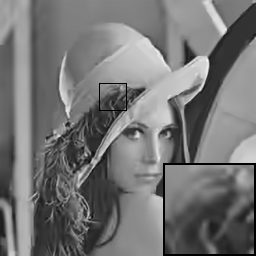}
\begin{scriptsize}
\centering
\end{scriptsize}
\end{minipage}
\\
\vskip -0.42 cm \hskip 0.12 cm \begin{tiny}Ground Truth \quad \quad \quad \quad \quad \quad \quad \quad \quad \quad \quad \ LRS-H \quad \quad \quad \quad \quad \quad \quad \quad \quad \quad \quad \quad \quad DLAMP-CS \quad \quad \quad \quad \quad \quad \quad \quad \quad \quad \ \quad CSCNet-H\end{tiny}
\vskip -0.1in
\caption{Visual quality comparisons between the proposed method and other CS-based image coding schemes in terms of Bpp=0.2 on image \emph{Lenna}.}
\label{Fig:a6}
\vskip -0.08in
\end{figure*}

\subsection{Comparisons with State-of-the-art Methods}

To evaluate the performance of the proposed CS coding framework, we conduct the experimental comparisons by three aspects: comparisons with the image CS coding (CSC) algorithms, the CS-based image coding (CSBC) schemes, and the existing image compression standards. Firstly, for the CSC method, three representative algorithms are considered, including two optimization-based CS coding methods SDPC (spatially directional predictive coding)~\cite{Zhang2013Spatially} and LSMM (local structural measurement matrix)~\cite{gao2015block} as well as one deep network-based CS coding scheme DQBCS (deep quantization block-based compressed sensing coding)~\cite{Cui2018An}. Secondly, for the CSBC method, two representative schemes are considered, including one optimization-based coding method LRS~\cite{Liu2016Compressive} and one deep network-based coding scheme DLAMP-CS~\cite{chen2019compressive}. Thirdly, for the image compression standards, two existing efficient image coding standards are concerned, i.e., JPEG2000$\footnote{http://www.openjpeg.org/}$ and BPG$\footnote{https://bellard.org/bpg/}$ (Based on HEVC-Intra). For the experiments of the proposed CSCNet, JPEG2000 and BPG are used as the third-party codecs to compress the produced measurements for producing two model variations CSCNet-J and CSCNet-H. Specifically, in the training process, the generated measurements are compressed using these two image codecs (JPEG2000, BPG), and in the testing process, the corresponding reconstruction results at different bit rates are obtained by adjusting the quantization parameter (BPG) or compression ratio (JPEG2000).

For testing data, we carry out extensive experiments on 8 representative test images used in work~\cite{chen2019compressive} (shown in Fig.~\ref{Fig:a1}), and the size of these images is 256$\times$256. Besides, several benchmark datasets are also tested in our experiments, such as Set11~\cite{zhang2018ista}. In terms of the experimental comparisons, the results of DLAMP-CS are obtained from~\cite{chen2019compressive}. For the other methods, we produce the experimental results by running their source codes downloaded from the authors' websites and we evaluate the coding performance with two widely used quality evaluation metrics: PSNR and SSIM in terms of different sampling ratios and bit rates.

\begin{figure*}[t]
\hspace{-0.1in}
\begin{minipage}[t]{0.17\textwidth}
\centering
\includegraphics[width=0.99in]{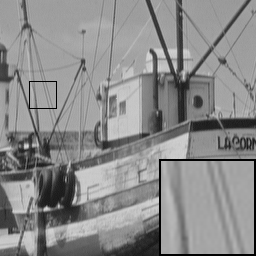}
\begin{tiny}
\centering

\end{tiny}
\end{minipage}
\hskip 0.3 cm
\begin{minipage}[t]{0.17\textwidth}
\centering
\includegraphics[width=0.99in]{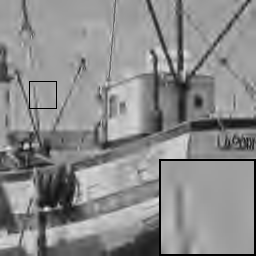}
\begin{scriptsize}
\centering
\end{scriptsize}
\end{minipage}
\hskip 0.3 cm
\begin{minipage}[t]{0.17\textwidth}
\centering
\includegraphics[width=0.99in]{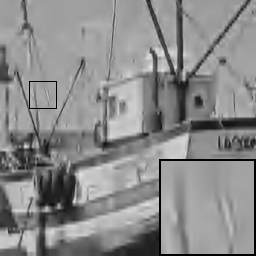}
\begin{scriptsize}
\centering
\end{scriptsize}
\end{minipage}
\hskip 0.3 cm
\begin{minipage}[t]{0.17\textwidth}
\centering
\includegraphics[width=0.99in]{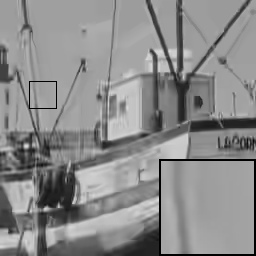}
\begin{scriptsize}
\centering
\end{scriptsize}
\end{minipage}
\hskip 0.3 cm
\begin{minipage}[t]{0.17\textwidth}
\centering
\includegraphics[width=0.99in]{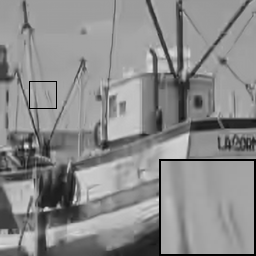}
\begin{scriptsize}
\centering
\end{scriptsize}
\end{minipage}
\\
\vskip -0.42 cm \hskip -0.2 cm \begin{tiny}Ground Truth \quad \quad \quad \quad \quad \quad \quad \quad \ JPEG2000 \quad \quad \quad \quad \quad \quad \quad \quad \quad \  CSCNet-J \quad \quad \quad \quad \quad \quad \quad \quad\ \quad BPG \quad \quad \quad \quad \quad \quad \quad \quad \ \quad CSCNet-H\end{tiny}

\vskip -0.07in
\vspace{-0.05in} \caption{Visual quality comparisons between the proposed method and two third-party image coding schemes in terms of Bpp=0.2 on image \emph{Boats}.}
\label{Fig:a11}
\vspace{-0.07in}
\end{figure*}

\begin{figure*}[t]
\hspace{-0.11in}
\begin{minipage}[t]{0.17\textwidth}
\centering
\includegraphics[width=0.99in]{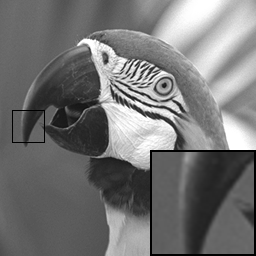}
\begin{tiny}
\centering
\end{tiny}
\end{minipage}
\hskip 0.3 cm
\begin{minipage}[t]{0.17\textwidth}
\centering
\includegraphics[width=0.99in]{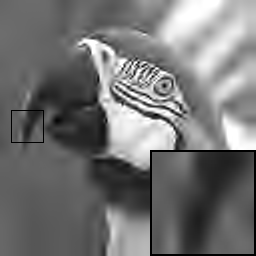}
\begin{scriptsize}
\centering
\end{scriptsize}
\end{minipage}
\hskip 0.3 cm
\begin{minipage}[t]{0.17\textwidth}
\centering
\includegraphics[width=0.99in]{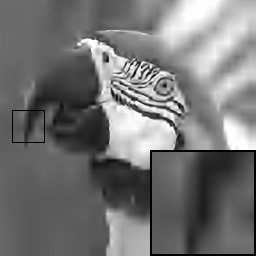}
\begin{scriptsize}
\centering
\end{scriptsize}
\end{minipage}
\hskip 0.3 cm
\begin{minipage}[t]{0.17\textwidth}
\centering
\includegraphics[width=0.99in]{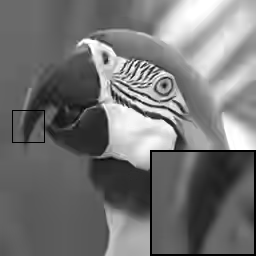}
\begin{scriptsize}
\centering
\end{scriptsize}
\end{minipage}
\hskip 0.3 cm
\begin{minipage}[t]{0.17\textwidth}
\centering
\includegraphics[width=0.99in]{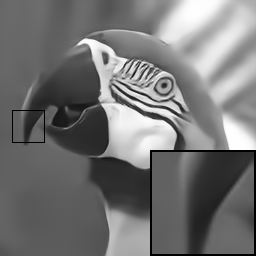}
\begin{scriptsize}
\centering
\end{scriptsize}
\end{minipage}
\\
\vskip -0.42 cm \hskip -0.2 cm \begin{tiny}Ground Truth \quad \quad \quad \quad \quad \quad \quad \quad \ JPEG2000 \quad \quad \quad \quad \quad \quad \quad \quad \quad \  CSCNet-J \quad \quad \quad \quad \quad \quad \quad \quad\ \quad BPG \quad \quad \quad \quad \quad \quad \quad \quad \ \quad CSCNet-H\end{tiny}

\vskip -0.1in
\caption{Visual quality comparisons between the proposed method and two third-party image coding schemes in terms of Bpp=0.1 on image \emph{Parrot}.}
\label{Fig:a12}
\vskip -0.15in
\end{figure*}

\subsubsection{Comparisons with CSC Method}

In this subsection, we compare the proposed CSCNet with other three image CS coding methods, i.e., SDPC, LSMM and DQBCS. For these three compared CS coding schemes, three sampling ratios (R), i.e., 0.1, 0.2 and 0.3, are mainly concerned. In terms of the proposed CSCNet, we use the image codec BPG to compress the produced measurements (dubbed CSCNet-H). The comparison results in terms of PSNR and SSIM are shown in Tables~\ref{Tab:a1} and~\ref{Tab:a2} respectively, from which we can get that our framework performs much better than other image CS coding methods. Specifically, for the optimization-based CS coding methods SDPC and LSMM, CSCNet-H achieves more than 1dB, 2dB and 3dB gains on average in PSNR for the given three sampling ratios. For the deep network-based method DQBCS, the proposed method achieves more than 0.8dB, 1dB and 2dB gains (PSNR) on average at the given three sampling ratios. The visual comparisons are shown in Figs.~\ref{Fig:a111} and~\ref{Fig:a222}, from which we observe that CSCNet-H is capable of preserving more structural details compared to the other image CS coding methods.

\subsubsection{Comparisons with CSBC Method}

In this subsection, we compare the proposed CSCNet with other two CS-based image coding methods, i.e., LRS and DLAMP-CS. Specifically, for LRS and the proposed CSCNet, the produced measurements are coded by JPEG2000 and BPG respectively, dubbed LRS-J, LRS-H and CSCNet-J, CSCNet-H. For LRS, its sampling ratio is set to 0.25. For the proposed CSCNet, the sampling ratios are set to 0.25 and 0.50. The quantitative results are shown in Table~\ref{Tab:a3}, from which we can get that the proposed CSCNet performs much better than LRS on the same sampling ratio (R=0.25). More specifically, for the image codec JPEG2000, CSCNet-J achieves on average 0.39dB, 0.60dB, 0.50dB, 0.42dB, 0.34dB and 0.015, 0.026, 0.026, 0.027, 0.023 gains in PSNR and SSIM for the given Bpps compared against LRS-J. For the image codec BPG, CSCNet-H achieves on average 0.82dB, 1.36dB, 1.77dB, 1.43dB, 1.32dB and 0.020, 0.034, 0.042, 0.028, 0.021 gains in PSNR and SSIM compared against LRS-H. For DLAMP-CS, the experimental results show that the proposed framework gets more higher coding efficiency at most cases. More specifically, for Bpp from 0.1 to 0.4, the proposed framework achieves on average 1.93dB, 0.64dB, 0.40dB, 0.09dB and 0.077, 0.035, 0.024, 0.020 gains in PSNR and SSIM. Specially, when Bpp is 0.5, the proposed method is inferior compared with DLAMP-CS in PSNR,  but is better in terms of SSIM. In addition, the experimental results show that the proposed CSCNet can obtain better subjective quality. The visual comparisons are shown in Fig.~\ref{Fig:a6}, from which we observe that the proposed CSCNet is capable of preserving more structural details compared to the other CS-based image coding methods.

\begin{table*}[t]
\renewcommand\arraystretch{0.9}
\centering
\caption{The average running time comparisons for different Compressed Sensing Coding schemes in terms of various bit rates.}
\vspace{-0.15in}
\label{Tab:a4}
\begin{threeparttable}

\begin{tabular}{p{2.25cm}<{\centering} | p{0.92cm}<{\centering} p{1.16cm}<{\centering} | p{0.92cm}<{\centering} p{1.16cm}<{\centering} | p{0.92cm}<{\centering} p{1.16cm}<{\centering} | p{0.92cm}<{\centering} p{1.16cm}<{\centering}}
\toprule
\multirow{2}*{\small{Method}} & \multicolumn{2}{c|}{\small{Bpp=}0.1} & \multicolumn{2}{c|}{\small{Bpp=}0.3} & \multicolumn{2}{c|}{\small{Bpp=}0.5} & \multicolumn{2}{c}{\small{Average}}\\
\cline{2-9}
&\small{Encoder}&\small{Decoder}&\small{Encoder}&\small{Decoder}&\small{Encoder}&\small{Decoder}&\small{Encoder}&\small{Decoder}\\
\midrule

\small{DQBCS}&0.0054&0.0268&0.0068&0.0344&0.0084&0.0426&0.0069&0.0346\\

\small{DLAMP-CS}&0.4567&--&0.4838&--&0.5201&--&0.4869&--\\
\hline

\small{JPEG2000}&0.0361&0.0096&0.0382&0.0117&0.0397&0.0132&0.0380&0.0115\\
\small{BPG}&0.2487&0.0763&0.2585&0.0790&0.2749&0.0860&0.2607&0.0804\\

\hline

\hline

\small{CSCNet-J-0.25}&0.0128&0.0406&0.0156&0.0450&0.0189&0.0497&0.0158&0.0451\\
\small{CSCNet-J-0.50}&0.0198&0.0347&0.0254&0.0395&0.0303&0.0439&0.0252&0.0394\\

\small{CSCNet-H-0.25}&0.1069&0.0745&0.1254&0.0804&0.1401&0.0894&0.1241&0.0814\\
\small{CSCNet-H-0.50}&0.1626&0.0606&0.1680&0.0655&0.1758&0.0704&0.1688&0.0655\\

\bottomrule

\end{tabular}

\begin{tablenotes}
\item[a] \small{The traditional image codecs JPEG2000 and BPG are implemented on CPU.}
\item[b] \small{The deep network-based schemes DQBCS, DLAMP-CS and the proposed CSCNet are implemented on GPU.}
\end{tablenotes}
\end{threeparttable}

\vspace{-0.24in}
\end{table*}

\subsubsection{Comparisons with Image Compression Standards}

In order to give the reader a sense of CSCNet compression performance, we also compare our framework with the existing image compression standards, including JPEG2000 and BPG (HEVC-Intra based image codec). Table~\ref{Tab:ccc} presents the comparisons against these two existing image coding standards. Specifically, for JPEG2000, the proposed variation based on JPEG2000 (CSCNet-J) can achieve better coding performance at most cases. More specifically, for Bpp from 0.1 to 0.4, the proposed framework (CSCNet-J-0.50) achieves on average 1.26dB, 0.94dB, 0.66dB, 0.08dB and 0.046, 0.030, 0.021, 0.016 gains in PSNR and SSIM compared with the image codec JPEG2000. Specially, when Bpp is 0.5, the proposed method is inferior compared with JPEG2000 in PSNR, but is better in terms of SSIM. For BPG, when Bpp $>$ 0.1, the PSNR of the proposed framework (CSCNet-H-0.50) cannot reach the coding performance of BPG, but the SSIM of our method is better in most cases (especially at low bit rates). The visual subjective comparisons are shown in Figs.~\ref{Fig:a11} and~\ref{Fig:a12}, which reveal that the proposed framework preserves more high-frequency information and retains sharper edges in the reconstructed images.

In fact, it is somewhat unreasonable to directly compare the proposed CS coding method with these traditional image compression standards, because these traditional image codecs take the original images or videos sampled according to the Nyquist sampling theorem as input. In contrast, the proposed method is based on the CS sampling system, in which the sampled measurements, instead of the original images (or videos), are encoded directly. In Table~\ref{Tab:ccc}, the comparison results show that the proposed CS coding method is inferior to the traditional image compression standards JPEG2000 and BPG in certain bitrates, which verifies that our proposed CS coding method cannot completely surpass the traditional image codecs. The purpose of the comparisons is to give the reader comparative levels of compression that can be expected in the future.

\subsubsection{Running Speed Comparisons}

To ensure the fairness of the comparisons, we evaluate the runtime on the same platform with 3.30 GHz Intel i7 CPU (32G RAM) plus NVIDIA GTX 1080Ti GPU (11G Memory). Table~\ref{Tab:a4} shows the average running time (in second) of different methods in terms of various compression rates on the 8 test images. For the compared methods, two deep learning based CS coding schemes DQBCS and DLAMP-CS as well as two traditional image codecs JPEG2000 and BPG are concerned. Besides, it is worth noting that the encoding runtime of DLAMP-CS is copied from~\cite{chen2019compressive}. The running speed comparison results show that the encoder of our preliminary version DQBCS has the fastest running speed compared with other methods. In addition, the encoder of the proposed CSCNet is faster than the corresponding traditional image codecs and the deep learning based CS coding scheme DLAMP-CS. For the decoder time, the proposed CSCNet slightly increases compared with the corresponding traditional image codecs. Because the optimization based CS coding schemes SDPC, LSMM and the CS-based image coding method LRS usually need hundreds of iterations to solve an inverse problem, which usually results in high computational cost. Thus, we do not show the runtime of these optimization-based coding methods.

\begin{table}[t]
\renewcommand\arraystretch{0.85}
\vskip -0.02in
\centering
\caption{The experimental results of the contributions for different functional modules in terms of various bit rates on dataset Set11. The comparison results are based on CSCNet-H-0.25.}
\label{Tab:a5}
\vspace{-0.15in}
\begin{tabular}{p{1.20cm}<{\centering} || p{1.1cm}<{\centering} p{1.1cm}<{\centering} | p{1.1cm}<{\centering}  p{1.1cm}<{\centering} | p{1.1cm}<{\centering} p{1.1cm}<{\centering} | p{1.1cm}<{\centering}  p{1.1cm}<{\centering}}

\toprule
\multirow{2}*{\small{Bpp}} & \multicolumn{2}{c|}{\small{Sampling}} & \multicolumn{2}{c|}{\small{Coding}} & \multicolumn{2}{c|}{\small{Reconstruction}} & \multicolumn{2}{c}{\small{CSCNet-H}}\\
\cline{2-9}
&\small{PSNR}&\small{SSIM}&\small{PSNR}&\small{SSIM}&\small{PSNR}&\small{SSIM}&\small{PSNR}&\small{SSIM}\\
\midrule

0.1&26.89&\textbf{0.792}&25.16&0.751&26.71&0.789&\textbf{26.90}&\textbf{0.792}\\
0.2&29.36&\textbf{0.866}&27.04&0.829&29.23&0.862&\textbf{29.39}&\textbf{0.866}\\
0.3&30.55&0.896&27.62&0.863&30.36&0.895&\textbf{30.59}&\textbf{0.897}\\
0.4&31.20&0.912&28.17&0.889&31.05&0.911&\textbf{31.25}&\textbf{0.913}\\
0.5&31.55&0.920&28.55&0.897&31.38&0.917&\textbf{31.61}&\textbf{0.921}\\
\midrule
\small{Avg}&29.91&0.877&27.31&0.846&29.75&0.875&\textbf{29.95}&\textbf{0.878}\\

\bottomrule

\end{tabular}
\vspace{-0.14in}
\end{table}

\subsection{Ablation Studies and Discussions}
\label{section:a5}

In this section, the contributions of each functional modules are mainly analyzed. Specifically, we first discuss the effect of the proposed learning strategy by comparing our learned local matrix against the hand-designed version. Subsequently, more analysis in terms of the measurement coding is provided. Finally, the efficiency of the proposed Laplacian pyramid reconstruction network is analyzed.

To verify the efficiency of the proposed learned local structural sampling matrix, we compare our learned local structural sampling matrix with the manually designed random local matrix (LSMM) in~\cite{gao2015block}. Specifically, in order to ensure the fairness and simplicity of comparison, we first use the LSMM to replace the proposed learned local matrix. Then, we fix the sampling matrix and retrain the model with the same training configuration. Table~\ref{Tab:a5} shows the performance comparison between the proposed learned local matrix and the manually designed local matrix (LSMM) in terms of various compression rates on the dataset Set11~\cite{zhang2018ista}, from which we can observe that the proposed learned local structural sampling matrix achieves on average 0.04dB and 0.001 gains in PSNR and SSIM compared against LSMM. Since the measurements sampled by the global random sampling matrix (such as GRM) cannot be directly compressed by the third-party image codecs, we cannot provide the comparisons with GRM.

For the measurement coding of LSMM, the prediction-based coding strategy is utilized, in which each measurement is predicted by the adjacent measurement coefficient to produce the measurement residuals, and then the entropy of the residuals is estimated to measure the coding efficiency. In contrast, the measurements of the proposed method CSCNet are coded by the third-party image codec. To ensure the fairness of comparison, we use the prediction-based measurement coding strategy of LSMM to compress the produced measurements of the proposed CS coding framework and the comparison results are shown in Table~\ref{Tab:a5}, from which we can obtain that the proposed measurement coding method (Using third-party image codec to compress measurements) can obtain more than 2.5dB gain in PSNR and 0.030 gain in SSIM on average. 

For image reconstruction module of our proposed CSCNet, a new convolutional Laplacian pyramid reconstruction network is proposed to reconstruct the target image from the quantized measurement domain to the image domain. In order to verify the efficiency of the proposed Laplacian pyramid network, we use the single scale network architecture of our early work DQBCS~\cite{Cui2018An} to reconstruct the target image and keep the other functional modules unchanged. The comparison results are shown in Table~\ref{Tab:a5}, from which we can observe that the proposed convolutional Laplacian pyramid network achieves better reconstruction quality.

From Table~\ref{Tab:a5}, the gain of our proposed local sampling matrix is smaller than the other functional modules (measurement coding and pyramid reconstruction network), the possible reason is analyzed below: in our proposed local matrix, only the elements in the local window are jointly optimized with other functional modules, which signifies there are very few learnable parameters in the proposed local structural sampling matrix. Furthermore, the constraint in Eq.~\ref{Eq:a6} further weakens the optimization space of the proposed sampling matrix. As mentioned above, the gain of the proposed local structural sampling matrix is severely limited. For the measurement coding module, compared to the manually designed measurement coding tools, the existing third-party image codecs often achieve greater gains by efficiently exploiting the correlations between the adjacent measurements. On the other hand, compared to the single scale network structure~\cite{Cui2018An}, the proposed Laplacian pyramid reconstruction network can efficiently explore and integrate richer deep features of different scales, thereby achieving better reconstruction performance.

As shown in our experiments, the proposed CS coding framework achieves better coding performance compared with the existing CS-related coding methods. However, there still exist the following two limitations: \textbf{1)} the quantization operator in existing third-party image coding algorithms is usually non-differentiable, which obviously affects the gradient backpropagation in the training process of our proposed CS coding network. In our framework, we roughly ignore the quantization loss in the training process, and approximately regard the derivative of the quantization operator as 1. In fact, the above approximation of derivatives is inaccurate, which mostly affects the computational accuracy of gradients during the training process, thus limiting the coding efficiency of the proposed CS coding model. \textbf{2)} As shown in Table 4, the proposed CS coding framework based on JPEG2000 is able to achieve better coding performance compared with the corresponding third-party image coding scheme. However, based on BPG (HEVC-Intra), the proposed framework typically cannot reach the performance of the corresponding third-party image coding methods under the evaluation criterion of PSNR especially at high bitrates. In the future, we will focus on the above issues to further improve CS coding performance.

\section{Conclusion}
\label{section:a6}

In this paper, a new deep network-based image compressed sensing coding framework using local structural sampling is proposed, in which three functional modules are included, i.e., local structural sampling, measurement coding and Laplacian pyramid reconstruction. In the proposed framework, the image is first sampled by using a local structural sampling matrix, after sampling the measurements with high correlations are generated. Then, the measurement coding module aims to compress the produced measurements into bitstreams. At last, a convolutional Laplacian pyramid architecture is proposed to reconstruct the target image from the quantized measurement domain to the image domain. Besides, it is worth noting that the proposed framework can be trained in an end-to-end fashion, which facilitates the communication between different modules for better coding performance. Experimental results show that the proposed framework outperforms the other state-of-the-art image CS coding (CSC) methods and CS-based image coding (CSBC) methods.

\section{ACKNOWLEDGMENTS}

This work was supported in part by the National Key R\&D Program of China (2021YFF0900500), and the National Natural Science Foundation of China (NSFC) under grants 62302128, 62272128.

\bibliographystyle{ACM-Reference-Format}
\bibliography{sample-base}


\begin{thebibliography}{70}


\ifx \showCODEN    \undefined \def \showCODEN     #1{\unskip}     \fi
\ifx \showDOI      \undefined \def \showDOI       #1{#1}\fi
\ifx \showISBNx    \undefined \def \showISBNx     #1{\unskip}     \fi
\ifx \showISBNxiii \undefined \def \showISBNxiii  #1{\unskip}     \fi
\ifx \showISSN     \undefined \def \showISSN      #1{\unskip}     \fi
\ifx \showLCCN     \undefined \def \showLCCN      #1{\unskip}     \fi
\ifx \shownote     \undefined \def \shownote      #1{#1}          \fi
\ifx \showarticletitle \undefined \def \showarticletitle #1{#1}   \fi
\ifx \showURL      \undefined \def \showURL       {\relax}        \fi
\providecommand\bibfield[2]{#2}
\providecommand\bibinfo[2]{#2}
\providecommand\natexlab[1]{#1}
\providecommand\showeprint[2][]{arXiv:#2}

\bibitem[{Agustsson} et~al\mbox{.}(2019)]%
        {9010721}
\bibfield{author}{\bibinfo{person}{E. {Agustsson}}, \bibinfo{person}{M.
  {Tschannen}}, \bibinfo{person}{F. {Mentzer}}, \bibinfo{person}{R. {Timofte}},
  {and} \bibinfo{person}{L. {Van Gool}}.} \bibinfo{year}{2019}\natexlab{}.
\newblock \showarticletitle{Generative Adversarial Networks for Extreme Learned
  Image Compression}. In \bibinfo{booktitle}{\emph{2019 IEEE/CVF International
  Conference on Computer Vision (ICCV)}}. \bibinfo{pages}{221--231}.
\newblock
\urldef\tempurl%
\url{https://doi.org/10.1109/ICCV.2019.00031}
\showDOI{\tempurl}


\bibitem[Cand{\`e}s and Wakin(2008)]%
        {candes2008introduction}
\bibfield{author}{\bibinfo{person}{Emmanuel~J Cand{\`e}s} {and}
  \bibinfo{person}{Michael~B Wakin}.} \bibinfo{year}{2008}\natexlab{}.
\newblock \showarticletitle{An introduction to compressive sampling}.
\newblock \bibinfo{journal}{\emph{IEEE Signal Processing Magazine}}
  \bibinfo{volume}{25}, \bibinfo{number}{2} (\bibinfo{year}{2008}),
  \bibinfo{pages}{21--30}.
\newblock


\bibitem[Chen and Zhang(2022)]%
        {chen2022content}
\bibfield{author}{\bibinfo{person}{Bin Chen} {and} \bibinfo{person}{Jian
  Zhang}.} \bibinfo{year}{2022}\natexlab{}.
\newblock \showarticletitle{Content-aware scalable deep compressed sensing}.
\newblock \bibinfo{journal}{\emph{IEEE Transactions on Image Processing}}
  \bibinfo{volume}{31} (\bibinfo{year}{2022}), \bibinfo{pages}{5412--5426}.
\newblock


\bibitem[{Chen} et~al\mbox{.}(2021)]%
        {9359473}
\bibfield{author}{\bibinfo{person}{T. {Chen}}, \bibinfo{person}{H. {Liu}},
  \bibinfo{person}{Z. {Ma}}, \bibinfo{person}{Q. {Shen}}, \bibinfo{person}{X.
  {Cao}}, {and} \bibinfo{person}{Y. {Wang}}.} \bibinfo{year}{2021}\natexlab{}.
\newblock \showarticletitle{End-to-End Learnt Image Compression via Non-Local
  Attention Optimization and Improved Context Modeling}.
\newblock \bibinfo{journal}{\emph{IEEE Transactions on Image Processing}}
  \bibinfo{volume}{30} (\bibinfo{year}{2021}), \bibinfo{pages}{3179--3191}.
\newblock
\urldef\tempurl%
\url{https://doi.org/10.1109/TIP.2021.3058615}
\showDOI{\tempurl}


\bibitem[Chen et~al\mbox{.}(2018)]%
        {8110646}
\bibfield{author}{\bibinfo{person}{Zan Chen}, \bibinfo{person}{Xingsong Hou},
  \bibinfo{person}{Xueming Qian}, {and} \bibinfo{person}{Chen Gong}.}
  \bibinfo{year}{2018}\natexlab{}.
\newblock \showarticletitle{Efficient and Robust Image Coding and Transmission
  Based on Scrambled Block Compressive Sensing}.
\newblock \bibinfo{journal}{\emph{IEEE Transactions on Multimedia}}
  \bibinfo{volume}{20}, \bibinfo{number}{7} (\bibinfo{year}{2018}),
  \bibinfo{pages}{1610--1621}.
\newblock



\bibitem[{Chen} et~al\mbox{.}(2020)]%
        {chen2019compressive}
\bibfield{author}{\bibinfo{person}{Z. {Chen}}, \bibinfo{person}{X. {Hou}},
  \bibinfo{person}{L. {Shao}}, \bibinfo{person}{C. {Gong}}, \bibinfo{person}{X.
  {Qian}}, \bibinfo{person}{Y. {Huang}}, {and} \bibinfo{person}{S. {Wang}}.}
  \bibinfo{year}{2020}\natexlab{}.
\newblock \showarticletitle{Compressive Sensing Multi-Layer Residual
  Coefficients for Image Coding}.
\newblock \bibinfo{journal}{\emph{IEEE Transactions on Circuits and Systems for
  Video Technology}} \bibinfo{volume}{30}, \bibinfo{number}{4}
  (\bibinfo{year}{2020}), \bibinfo{pages}{1109--1120}.
\newblock
\urldef\tempurl%
\url{https://doi.org/10.1109/TCSVT.2019.2898908}
\showDOI{\tempurl}


\bibitem[Cho and Yu(2020)]%
        {8902011}
\bibfield{author}{\bibinfo{person}{Wonwoo Cho} {and} \bibinfo{person}{Nam~Yul
  Yu}.} \bibinfo{year}{2020}\natexlab{}.
\newblock \showarticletitle{Secure and Efficient Compressed Sensing-Based
  Encryption With Sparse Matrices}.
\newblock \bibinfo{journal}{\emph{IEEE Transactions on Information Forensics
  and Security}}  \bibinfo{volume}{15} (\bibinfo{year}{2020}),
  \bibinfo{pages}{1999--2011}.
\newblock
\urldef\tempurl%
\url{https://doi.org/10.1109/TIFS.2019.2953383}
\showDOI{\tempurl}


\bibitem[Cui et~al\mbox{.}(2018)]%
        {Cui2018An}
\bibfield{author}{\bibinfo{person}{Wenxue Cui}, \bibinfo{person}{Feng Jiang},
  \bibinfo{person}{Xinwei Gao}, \bibinfo{person}{Shengping Zhang}, {and}
  \bibinfo{person}{Debin Zhao}.} \bibinfo{year}{2018}\natexlab{}.
\newblock \showarticletitle{An Efficient Deep Quantized Compressed Sensing
  Coding Framework of Natural Images}.
\newblock \bibinfo{journal}{\emph{ACM International Conference on Multimedia
  (MM)}} (\bibinfo{year}{2018}), \bibinfo{pages}{1777--1785}.
\newblock
\urldef\tempurl%
\url{https://doi.org/10.1145/3240508.3240706}
\showDOI{\tempurl}


\bibitem[Daubechies et~al\mbox{.}(2004)]%
        {daubechies2004iterative}
\bibfield{author}{\bibinfo{person}{Ingrid Daubechies}, \bibinfo{person}{Michel
  Defrise}, {and} \bibinfo{person}{Christine De~Mol}.}
  \bibinfo{year}{2004}\natexlab{}.
\newblock \showarticletitle{An iterative thresholding algorithm for linear
  inverse problems with a sparsity constraint}.
\newblock \bibinfo{journal}{\emph{Communications on Pure and Applied
  Mathematics: A Journal Issued by the Courant Institute of Mathematical
  Sciences}} \bibinfo{volume}{57}, \bibinfo{number}{11} (\bibinfo{year}{2004}),
  \bibinfo{pages}{1413--1457}.
\newblock


\bibitem[{Dinh} and {Jeon}(2017)]%
        {7505983}
\bibfield{author}{\bibinfo{person}{K.~Q. {Dinh}} {and} \bibinfo{person}{B.
  {Jeon}}.} \bibinfo{year}{2017}\natexlab{}.
\newblock \showarticletitle{Iterative Weighted Recovery for Block-Based
  Compressive Sensing of Image/Video at a Low Subrate}.
\newblock \bibinfo{journal}{\emph{IEEE Transactions on Circuits and Systems for
  Video Technology}} \bibinfo{volume}{27}, \bibinfo{number}{11}
  (\bibinfo{year}{2017}), \bibinfo{pages}{2294--2308}.
\newblock
\urldef\tempurl%
\url{https://doi.org/10.1109/TCSVT.2016.2587398}
\showDOI{\tempurl}


\bibitem[Dinh et~al\mbox{.}(2013)]%
        {6738003}
\bibfield{author}{\bibinfo{person}{Khanh~Quoc Dinh}, \bibinfo{person}{Hiuk~Jae
  Shim}, {and} \bibinfo{person}{Byeungwoo Jeon}.}
  \bibinfo{year}{2013}\natexlab{}.
\newblock \showarticletitle{Measurement coding for compressive imaging using a
  structural measuremnet matrix}. In \bibinfo{booktitle}{\emph{2013 IEEE
  International Conference on Image Processing}}. \bibinfo{pages}{10--13}.
\newblock
\urldef\tempurl%
\url{https://doi.org/10.1109/ICIP.2013.6738003}
\showDOI{\tempurl}


\bibitem[Dong et~al\mbox{.}(2014)]%
        {dong2014compressive}
\bibfield{author}{\bibinfo{person}{Weisheng Dong}, \bibinfo{person}{Guangming
  Shi}, \bibinfo{person}{Xin Li}, \bibinfo{person}{Yi Ma}, {and}
  \bibinfo{person}{Feng Huang}.} \bibinfo{year}{2014}\natexlab{}.
\newblock \showarticletitle{Compressive sensing via nonlocal low-rank
  regularization}.
\newblock \bibinfo{journal}{\emph{IEEE Transactions on Image Processing (TIP)}}
  \bibinfo{volume}{23}, \bibinfo{number}{8} (\bibinfo{year}{2014}),
  \bibinfo{pages}{3618--3632}.
\newblock


\bibitem[Donoho(2006)]%
        {donoho2006compressed}
\bibfield{author}{\bibinfo{person}{David~L Donoho}.}
  \bibinfo{year}{2006}\natexlab{}.
\newblock \showarticletitle{Compressed sensing}.
\newblock \bibinfo{journal}{\emph{IEEE Transactions on Information Theory}}
  \bibinfo{volume}{52}, \bibinfo{number}{4} (\bibinfo{year}{2006}),
  \bibinfo{pages}{1289--1306}.
\newblock


\bibitem[Duarte et~al\mbox{.}(2008)]%
        {4472247}
\bibfield{author}{\bibinfo{person}{Marco~F. Duarte}, \bibinfo{person}{Mark~A.
  Davenport}, \bibinfo{person}{Dharmpal Takhar}, \bibinfo{person}{Jason~N.
  Laska}, \bibinfo{person}{Ting Sun}, \bibinfo{person}{Kevin~F. Kelly}, {and}
  \bibinfo{person}{Richard~G. Baraniuk}.} \bibinfo{year}{2008}\natexlab{}.
\newblock \showarticletitle{Single-pixel imaging via compressive sampling}.
\newblock \bibinfo{journal}{\emph{IEEE Signal Processing Magazine}}
  \bibinfo{volume}{25}, \bibinfo{number}{2} (\bibinfo{year}{2008}),
  \bibinfo{pages}{83--91}.
\newblock
\urldef\tempurl%
\url{https://doi.org/10.1109/MSP.2007.914730}
\showDOI{\tempurl}


\bibitem[Ebrahim and Chia(2015)]%
        {10.1145/2818712}
\bibfield{author}{\bibinfo{person}{Mansoor Ebrahim} {and}
  \bibinfo{person}{Wai~Chong Chia}.} \bibinfo{year}{2015}\natexlab{}.
\newblock \showarticletitle{Multiview Image Block Compressive Sensing with
  Joint Multiphase Decoding for Visual Sensor Network}.
\newblock \bibinfo{journal}{\emph{ACM Trans. Multimedia Comput. Commun. Appl.}}
  \bibinfo{volume}{12}, \bibinfo{number}{2}, Article \bibinfo{articleno}{30}
  (\bibinfo{date}{oct} \bibinfo{year}{2015}), \bibinfo{numpages}{23}~pages.
\newblock
\showISSN{1551-6857}
\urldef\tempurl%
\url{https://doi.org/10.1145/2818712}
\showDOI{\tempurl}


\bibitem[Everingham and Winn(2011)]%
        {everingham2011pascal}
\bibfield{author}{\bibinfo{person}{Mark Everingham} {and} \bibinfo{person}{John
  Winn}.} \bibinfo{year}{2011}\natexlab{}.
\newblock \bibinfo{title}{The PASCAL Visual Object Classes Challenge 2012
  ({VOC}2012) Development Kit}.
\newblock
\newblock


\bibitem[Gan(2007)]%
        {gan2007block}
\bibfield{author}{\bibinfo{person}{Lu Gan}.} \bibinfo{year}{2007}\natexlab{}.
\newblock \showarticletitle{Block compressed sensing of natural images}.
\newblock \bibinfo{journal}{\emph{Proceedings of the international conference
  on digital signal processing}} (\bibinfo{year}{2007}),
  \bibinfo{pages}{403--406}.
\newblock


\bibitem[Gao et~al\mbox{.}(2015)]%
        {gao2015block}
\bibfield{author}{\bibinfo{person}{Xinwei Gao}, \bibinfo{person}{Jian Zhang},
  \bibinfo{person}{Wenbin Che}, \bibinfo{person}{Xiaopeng Fan}, {and}
  \bibinfo{person}{Debin Zhao}.} \bibinfo{year}{2015}\natexlab{}.
\newblock \showarticletitle{Block-based compressive sensing coding of natural
  images by local structural measurement matrix}.
\newblock \bibinfo{journal}{\emph{IEEE Data Compression Conference (DCC)}}
  (\bibinfo{year}{2015}), \bibinfo{pages}{133--142}.
\newblock


\bibitem[Gao et~al\mbox{.}(2023)]%
        {10032194}
\bibfield{author}{\bibinfo{person}{Zhifan Gao}, \bibinfo{person}{Yifeng Guo},
  \bibinfo{person}{Jiajing Zhang}, \bibinfo{person}{Tieyong Zeng}, {and}
  \bibinfo{person}{Guang Yang}.} \bibinfo{year}{2023}\natexlab{}.
\newblock \showarticletitle{Hierarchical Perception Adversarial Learning
  Framework for Compressed Sensing MRI}.
\newblock \bibinfo{journal}{\emph{IEEE Transactions on Medical Imaging}}
  (\bibinfo{year}{2023}), \bibinfo{pages}{1--1}.
\newblock
\urldef\tempurl%
\url{https://doi.org/10.1109/TMI.2023.3240862}
\showDOI{\tempurl}


\bibitem[He et~al\mbox{.}(2015)]%
        {he2015delving}
\bibfield{author}{\bibinfo{person}{Kaiming He}, \bibinfo{person}{Xiangyu
  Zhang}, \bibinfo{person}{Shaoqing Ren}, {and} \bibinfo{person}{Jian Sun}.}
  \bibinfo{year}{2015}\natexlab{}.
\newblock \showarticletitle{Delving deep into rectifiers: Surpassing
  human-level performance on imagenet classification}.
\newblock \bibinfo{journal}{\emph{IEEE International Conference on Computer
  Vision (ICCV)}} (\bibinfo{year}{2015}), \bibinfo{pages}{1026--1034}.
\newblock


\bibitem[Huang et~al\mbox{.}(2013)]%
        {6738433}
\bibfield{author}{\bibinfo{person}{Gang Huang}, \bibinfo{person}{Hong Jiang},
  \bibinfo{person}{Kim Matthews}, {and} \bibinfo{person}{Paul Wilford}.}
  \bibinfo{year}{2013}\natexlab{}.
\newblock \showarticletitle{Lensless imaging by compressive sensing}. In
  \bibinfo{booktitle}{\emph{2013 IEEE International Conference on Image
  Processing}}. \bibinfo{pages}{2101--2105}.
\newblock
\urldef\tempurl%
\url{https://doi.org/10.1109/ICIP.2013.6738433}
\showDOI{\tempurl}


\bibitem[Jacques et~al\mbox{.}(2011)]%
        {Jacques2011Dequantizing}
\bibfield{author}{\bibinfo{person}{Laurent Jacques}, \bibinfo{person}{David~K
  Hammond}, {and} \bibinfo{person}{Jalal~M Fadili}.}
  \bibinfo{year}{2011}\natexlab{}.
\newblock \showarticletitle{Dequantizing compressed sensing: When oversampling
  and non-gaussian constraints combine}.
\newblock \bibinfo{journal}{\emph{IEEE Transactions on Information Theory}}
  \bibinfo{volume}{57}, \bibinfo{number}{1} (\bibinfo{year}{2011}),
  \bibinfo{pages}{559--571}.
\newblock


\bibitem[{Jia} et~al\mbox{.}(2019)]%
        {8574895}
\bibfield{author}{\bibinfo{person}{C. {Jia}}, \bibinfo{person}{X. {Zhang}},
  \bibinfo{person}{S. {Wang}}, \bibinfo{person}{S. {Wang}}, {and}
  \bibinfo{person}{S. {Ma}}.} \bibinfo{year}{2019}\natexlab{}.
\newblock \showarticletitle{Light Field Image Compression Using Generative
  Adversarial Network-Based View Synthesis}.
\newblock \bibinfo{journal}{\emph{IEEE Journal on Emerging and Selected Topics
  in Circuits and Systems}} \bibinfo{volume}{9}, \bibinfo{number}{1}
  (\bibinfo{year}{2019}), \bibinfo{pages}{177--189}.
\newblock
\urldef\tempurl%
\url{https://doi.org/10.1109/JETCAS.2018.2886642}
\showDOI{\tempurl}


\bibitem[{Jiang} et~al\mbox{.}(2018)]%
        {7999241}
\bibfield{author}{\bibinfo{person}{F. {Jiang}}, \bibinfo{person}{W. {Tao}},
  \bibinfo{person}{S. {Liu}}, \bibinfo{person}{J. {Ren}}, \bibinfo{person}{X.
  {Guo}}, {and} \bibinfo{person}{D. {Zhao}}.} \bibinfo{year}{2018}\natexlab{}.
\newblock \showarticletitle{An End-to-End Compression Framework Based on
  Convolutional Neural Networks}.
\newblock \bibinfo{journal}{\emph{IEEE Transactions on Circuits and Systems for
  Video Technology}} \bibinfo{volume}{28}, \bibinfo{number}{10}
  (\bibinfo{year}{2018}), \bibinfo{pages}{3007--3018}.
\newblock
\urldef\tempurl%
\url{https://doi.org/10.1109/TCSVT.2017.2734838}
\showDOI{\tempurl}


\bibitem[Li et~al\mbox{.}(2013)]%
        {li2013tval3}
\bibfield{author}{\bibinfo{person}{C Li}, \bibinfo{person}{W Yin}, {and}
  \bibinfo{person}{Y Zhang}.} \bibinfo{year}{2013}\natexlab{}.
\newblock \bibinfo{title}{Tval3: Tv minimization by augmented lagrangian and
  alternating direction agorithm 2009}.
\newblock
\newblock


\bibitem[{Li} et~al\mbox{.}(2020a)]%
        {9067005}
\bibfield{author}{\bibinfo{person}{M. {Li}}, \bibinfo{person}{K. {Ma}},
  \bibinfo{person}{J. {You}}, \bibinfo{person}{D. {Zhang}}, {and}
  \bibinfo{person}{W. {Zuo}}.} \bibinfo{year}{2020}\natexlab{a}.
\newblock \showarticletitle{Efficient and Effective Context-Based Convolutional
  Entropy Modeling for Image Compression}.
\newblock \bibinfo{journal}{\emph{IEEE Transactions on Image Processing}}
  \bibinfo{volume}{29} (\bibinfo{year}{2020}), \bibinfo{pages}{5900--5911}.
\newblock
\urldef\tempurl%
\url{https://doi.org/10.1109/TIP.2020.2985225}
\showDOI{\tempurl}


\bibitem[{Li} et~al\mbox{.}(2020b)]%
        {9050860}
\bibfield{author}{\bibinfo{person}{M. {Li}}, \bibinfo{person}{W. {Zuo}},
  \bibinfo{person}{S. {Gu}}, \bibinfo{person}{J. {You}}, {and}
  \bibinfo{person}{D. {Zhang}}.} \bibinfo{year}{2020}\natexlab{b}.
\newblock \showarticletitle{Learning Content-Weighted Deep Image Compression}.
\newblock \bibinfo{journal}{\emph{IEEE Transactions on Pattern Analysis and
  Machine Intelligence}} (\bibinfo{year}{2020}), \bibinfo{pages}{1--1}.
\newblock
\urldef\tempurl%
\url{https://doi.org/10.1109/TPAMI.2020.2983926}
\showDOI{\tempurl}


\bibitem[{Li} et~al\mbox{.}(2019)]%
        {8476610}
\bibfield{author}{\bibinfo{person}{Y. {Li}}, \bibinfo{person}{D. {Liu}},
  \bibinfo{person}{H. {Li}}, \bibinfo{person}{L. {Li}}, \bibinfo{person}{Z.
  {Li}}, {and} \bibinfo{person}{F. {Wu}}.} \bibinfo{year}{2019}\natexlab{}.
\newblock \showarticletitle{Learning a Convolutional Neural Network for Image
  Compact-Resolution}.
\newblock \bibinfo{journal}{\emph{IEEE Transactions on Image Processing}}
  \bibinfo{volume}{28}, \bibinfo{number}{3} (\bibinfo{year}{2019}),
  \bibinfo{pages}{1092--1107}.
\newblock
\urldef\tempurl%
\url{https://doi.org/10.1109/TIP.2018.2872876}
\showDOI{\tempurl}


\bibitem[{Li} et~al\mbox{.}(2018)]%
        {7982641}
\bibfield{author}{\bibinfo{person}{Y. {Li}}, \bibinfo{person}{D. {Liu}},
  \bibinfo{person}{H. {Li}}, \bibinfo{person}{L. {Li}}, \bibinfo{person}{F.
  {Wu}}, \bibinfo{person}{H. {Zhang}}, {and} \bibinfo{person}{H. {Yang}}.}
  \bibinfo{year}{2018}\natexlab{}.
\newblock \showarticletitle{Convolutional Neural Network-Based Block
  Up-Sampling for Intra Frame Coding}.
\newblock \bibinfo{journal}{\emph{IEEE Transactions on Circuits and Systems for
  Video Technology}} \bibinfo{volume}{28}, \bibinfo{number}{9}
  (\bibinfo{year}{2018}), \bibinfo{pages}{2316--2330}.
\newblock
\urldef\tempurl%
\url{https://doi.org/10.1109/TCSVT.2017.2727682}
\showDOI{\tempurl}


\bibitem[Li et~al\mbox{.}(2016)]%
        {li2016image}
\bibfield{author}{\bibinfo{person}{Yinghua Li}, \bibinfo{person}{Bin Song},
  \bibinfo{person}{Rong Cao}, \bibinfo{person}{Yue Zhang}, {and}
  \bibinfo{person}{Hao Qin}.} \bibinfo{year}{2016}\natexlab{}.
\newblock \showarticletitle{Image encryption based on compressive sensing and
  scrambled index for secure multimedia transmission}.
\newblock \bibinfo{journal}{\emph{ACM Transactions on Multimedia Computing,
  Communications, and Applications (TOMM)}} \bibinfo{volume}{12},
  \bibinfo{number}{4s} (\bibinfo{year}{2016}), \bibinfo{pages}{1--22}.
\newblock


\bibitem[{Lin} et~al\mbox{.}(2019)]%
        {8554306}
\bibfield{author}{\bibinfo{person}{J. {Lin}}, \bibinfo{person}{D. {Liu}},
  \bibinfo{person}{H. {Yang}}, \bibinfo{person}{H. {Li}}, {and}
  \bibinfo{person}{F. {Wu}}.} \bibinfo{year}{2019}\natexlab{}.
\newblock \showarticletitle{Convolutional Neural Network-Based Block
  Up-Sampling for {HEVC}}.
\newblock \bibinfo{journal}{\emph{IEEE Transactions on Circuits and Systems for
  Video Technology}} \bibinfo{volume}{29}, \bibinfo{number}{12}
  (\bibinfo{year}{2019}), \bibinfo{pages}{3701--3715}.
\newblock
\urldef\tempurl%
\url{https://doi.org/10.1109/TCSVT.2018.2884203}
\showDOI{\tempurl}


\bibitem[Liu et~al\mbox{.}(2016)]%
        {Liu2016Compressive}
\bibfield{author}{\bibinfo{person}{Xianming Liu}, \bibinfo{person}{Deming
  Zhai}, \bibinfo{person}{Jiantao Zhou}, \bibinfo{person}{Xinfeng Zhang},
  \bibinfo{person}{Debin Zhao}, {and} \bibinfo{person}{Wen Gao}.}
  \bibinfo{year}{2016}\natexlab{}.
\newblock \showarticletitle{Compressive Sampling-Based Image Coding for
  Resource-Deficient Visual Communication}.
\newblock \bibinfo{journal}{\emph{IEEE Transactions on Image Processing}}
  \bibinfo{volume}{25}, \bibinfo{number}{6} (\bibinfo{year}{2016}),
  \bibinfo{pages}{2844--2855}.
\newblock


\bibitem[Liu et~al\mbox{.}(2022)]%
        {liu2022deep}
\bibfield{author}{\bibinfo{person}{Zhi Liu}, \bibinfo{person}{Shuyuan Yang},
  \bibinfo{person}{Zhixi Feng}, \bibinfo{person}{Min Wang}, {and}
  \bibinfo{person}{Zhifan Yu}.} \bibinfo{year}{2022}\natexlab{}.
\newblock \showarticletitle{Deep Compressive Imaging With Meta-Learning}.
\newblock \bibinfo{journal}{\emph{IEEE Transactions on Instrumentation and
  Measurement}}  \bibinfo{volume}{72} (\bibinfo{year}{2022}),
  \bibinfo{pages}{1--9}.
\newblock


\bibitem[Lustig et~al\mbox{.}(2008)]%
        {Lustig2008Compressed}
\bibfield{author}{\bibinfo{person}{Michael Lustig}, \bibinfo{person}{David~L.
  Donoho}, \bibinfo{person}{Juan~M. Santos}, {and} \bibinfo{person}{John~M.
  Pauly}.} \bibinfo{year}{2008}\natexlab{}.
\newblock \showarticletitle{Compressed Sensing MRI}.
\newblock \bibinfo{journal}{\emph{IEEE Signal Processing Magazine}}
  \bibinfo{volume}{25}, \bibinfo{number}{2} (\bibinfo{year}{2008}),
  \bibinfo{pages}{72--82}.
\newblock


\bibitem[Mallat and Zhang(1993)]%
        {mallat1993matching}
\bibfield{author}{\bibinfo{person}{St{\'e}phane~G Mallat} {and}
  \bibinfo{person}{Zhifeng Zhang}.} \bibinfo{year}{1993}\natexlab{}.
\newblock \showarticletitle{Matching pursuits with time-frequency
  dictionaries}.
\newblock \bibinfo{journal}{\emph{IEEE Transactions on signal processing}}
  \bibinfo{volume}{41}, \bibinfo{number}{12} (\bibinfo{year}{1993}),
  \bibinfo{pages}{3397--3415}.
\newblock


\bibitem[Metzler et~al\mbox{.}(2016)]%
        {Metzler2016From}
\bibfield{author}{\bibinfo{person}{Christopher~A. Metzler},
  \bibinfo{person}{Arian Maleki}, {and} \bibinfo{person}{Richard~G. Baraniuk}.}
  \bibinfo{year}{2016}\natexlab{}.
\newblock \showarticletitle{From Denoising to Compressed Sensing}.
\newblock \bibinfo{journal}{\emph{IEEE Transactions on Information Theory}}
  \bibinfo{volume}{62}, \bibinfo{number}{9} (\bibinfo{year}{2016}),
  \bibinfo{pages}{5117--5144}.
\newblock


\bibitem[Mousavi et~al\mbox{.}(2015)]%
        {mousavi2015deep}
\bibfield{author}{\bibinfo{person}{Ali Mousavi}, \bibinfo{person}{Ankit~B
  Patel}, {and} \bibinfo{person}{Richard~G Baraniuk}.}
  \bibinfo{year}{2015}\natexlab{}.
\newblock \showarticletitle{A deep learning approach to structured signal
  recovery}.
\newblock \bibinfo{journal}{\emph{IEEE Allerton Conference on Communication,
  Control, and Computing (Allerton)}} (\bibinfo{year}{2015}),
  \bibinfo{pages}{1336--1343}.
\newblock


\bibitem[Mun and Fowler(2012)]%
        {Mun2012DPCM}
\bibfield{author}{\bibinfo{person}{Sungkwang Mun} {and}
  \bibinfo{person}{James~E Fowler}.} \bibinfo{year}{2012}\natexlab{}.
\newblock \showarticletitle{{DPCM} for quantized blockbased compressed sensing
  of images}.
\newblock \bibinfo{journal}{\emph{IEEE Signal Processing Conference}}
  (\bibinfo{year}{2012}), \bibinfo{pages}{1424--1428}.
\newblock


\bibitem[Shi et~al\mbox{.}(2019)]%
        {shi2019scalable}
\bibfield{author}{\bibinfo{person}{Wuzhen Shi}, \bibinfo{person}{Feng Jiang},
  \bibinfo{person}{Shaohui Liu}, {and} \bibinfo{person}{Debin Zhao}.}
  \bibinfo{year}{2019}\natexlab{}.
\newblock \showarticletitle{Scalable Convolutional Neural Network for Image
  Compressed Sensing}. In \bibinfo{booktitle}{\emph{Proceedings of the IEEE
  Conference on Computer Vision and Pattern Recognition}}.
  \bibinfo{pages}{12290--12299}.
\newblock


\bibitem[Shi et~al\mbox{.}(2020)]%
        {shi2019image}
\bibfield{author}{\bibinfo{person}{Wuzhen Shi}, \bibinfo{person}{Feng Jiang},
  \bibinfo{person}{Shaohui Liu}, {and} \bibinfo{person}{Debin Zhao}.}
  \bibinfo{year}{2020}\natexlab{}.
\newblock \showarticletitle{Image Compressed Sensing Using Convolutional Neural
  Network}.
\newblock \bibinfo{journal}{\emph{IEEE Transactions on Image Processing}}
  \bibinfo{volume}{29}, \bibinfo{number}{1} (\bibinfo{year}{2020}),
  \bibinfo{pages}{375--388}.
\newblock


\bibitem[{Shi} et~al\mbox{.}(2021)]%
        {9025255}
\bibfield{author}{\bibinfo{person}{W. {Shi}}, \bibinfo{person}{S. {Liu}},
  \bibinfo{person}{F. {Jiang}}, {and} \bibinfo{person}{D. {Zhao}}.}
  \bibinfo{year}{2021}\natexlab{}.
\newblock \showarticletitle{Video Compressed Sensing Using a Convolutional
  Neural Network}.
\newblock \bibinfo{journal}{\emph{IEEE Transactions on Circuits and Systems for
  Video Technology}} \bibinfo{volume}{31}, \bibinfo{number}{2}
  (\bibinfo{year}{2021}), \bibinfo{pages}{425--438}.
\newblock
\urldef\tempurl%
\url{https://doi.org/10.1109/TCSVT.2020.2978703}
\showDOI{\tempurl}


\bibitem[Shi et~al\mbox{.}(2023)]%
        {10049122}
\bibfield{author}{\bibinfo{person}{Wuzhen Shi}, \bibinfo{person}{Fei Tao},
  {and} \bibinfo{person}{Yang Wen}.} \bibinfo{year}{2023}\natexlab{}.
\newblock \showarticletitle{Structure-Aware Deep Networks and Pixel-Level
  Generative Adversarial Training for Single Image Super-Resolution}.
\newblock \bibinfo{journal}{\emph{IEEE Transactions on Instrumentation and
  Measurement}}  \bibinfo{volume}{72} (\bibinfo{year}{2023}),
  \bibinfo{pages}{1--14}.
\newblock
\urldef\tempurl%
\url{https://doi.org/10.1109/TIM.2023.3246523}
\showDOI{\tempurl}


\bibitem[Song et~al\mbox{.}(2021)]%
        {song2021memory}
\bibfield{author}{\bibinfo{person}{Jiechong Song}, \bibinfo{person}{Bin Chen},
  {and} \bibinfo{person}{Jian Zhang}.} \bibinfo{year}{2021}\natexlab{}.
\newblock \showarticletitle{Memory-augmented deep unfolding network for
  compressive sensing}. In \bibinfo{booktitle}{\emph{Proceedings of the 29th
  ACM International Conference on Multimedia}}. \bibinfo{pages}{4249--4258}.
\newblock


\bibitem[Sullivan et~al\mbox{.}(2012)]%
        {Sullivan2012Overview}
\bibfield{author}{\bibinfo{person}{Gary~J Sullivan}, \bibinfo{person}{Jens
  Ohm}, \bibinfo{person}{Woo~Jin Han}, {and} \bibinfo{person}{Thomas Wiegand}.}
  \bibinfo{year}{2012}\natexlab{}.
\newblock \showarticletitle{Overview of the high efficiency video coding
  ({HEVC}) standard}.
\newblock \bibinfo{journal}{\emph{IEEE Transactions on Circuits and Systems for
  Video Technology}} \bibinfo{volume}{22}, \bibinfo{number}{12}
  (\bibinfo{year}{2012}), \bibinfo{pages}{1649--1668}.
\newblock


\bibitem[Sun et~al\mbox{.}(2020)]%
        {9199540}
\bibfield{author}{\bibinfo{person}{Yubao Sun}, \bibinfo{person}{Jiwei Chen},
  \bibinfo{person}{Qingshan Liu}, \bibinfo{person}{Bo Liu}, {and}
  \bibinfo{person}{Guodong Guo}.} \bibinfo{year}{2020}\natexlab{}.
\newblock \showarticletitle{Dual-Path Attention Network for Compressed Sensing
  Image Reconstruction}.
\newblock \bibinfo{journal}{\emph{IEEE Transactions on Image Processing}}
  \bibinfo{volume}{29} (\bibinfo{year}{2020}), \bibinfo{pages}{9482--9495}.
\newblock
\urldef\tempurl%
\url{https://doi.org/10.1109/TIP.2020.3023629}
\showDOI{\tempurl}


\bibitem[Tang et~al\mbox{.}(2019)]%
        {tang2019feature}
\bibfield{author}{\bibinfo{person}{Chaoqing Tang}, \bibinfo{person}{Guiyun
  Tian}, \bibinfo{person}{Said Boussakta}, {and} \bibinfo{person}{Jianbo Wu}.}
  \bibinfo{year}{2019}\natexlab{}.
\newblock \showarticletitle{Feature-supervised compressed sensing for microwave
  imaging systems}.
\newblock \bibinfo{journal}{\emph{IEEE Transactions on Instrumentation and
  Measurement}} \bibinfo{volume}{69}, \bibinfo{number}{8}
  (\bibinfo{year}{2019}), \bibinfo{pages}{5287--5297}.
\newblock


\bibitem[Tran et~al\mbox{.}(2020)]%
        {9287074}
\bibfield{author}{\bibinfo{person}{Thuy T.~T. Tran}, \bibinfo{person}{Jirayu
  Peetakul}, \bibinfo{person}{Chi D.~K. Pham}, {and} \bibinfo{person}{Jinjia
  Zhou}.} \bibinfo{year}{2020}\natexlab{}.
\newblock \showarticletitle{Bi-directional intra prediction based measurement
  coding for compressive sensing images}. In \bibinfo{booktitle}{\emph{2020
  IEEE 22nd International Workshop on Multimedia Signal Processing (MMSP)}}.
  \bibinfo{pages}{1--6}.
\newblock
\urldef\tempurl%
\url{https://doi.org/10.1109/MMSP48831.2020.9287074}
\showDOI{\tempurl}


\bibitem[Tropp and Gilbert(2007)]%
        {tropp2007signal}
\bibfield{author}{\bibinfo{person}{Joel~A Tropp} {and} \bibinfo{person}{Anna~C
  Gilbert}.} \bibinfo{year}{2007}\natexlab{}.
\newblock \showarticletitle{Signal recovery from random measurements via
  orthogonal matching pursuit}.
\newblock \bibinfo{journal}{\emph{IEEE Transactions on information theory}}
  \bibinfo{volume}{53}, \bibinfo{number}{12} (\bibinfo{year}{2007}),
  \bibinfo{pages}{4655--4666}.
\newblock


\bibitem[Wan et~al\mbox{.}(2022)]%
        {9508849}
\bibfield{author}{\bibinfo{person}{Rentao Wan}, \bibinfo{person}{Jinjia Zhou},
  \bibinfo{person}{Bowen Huang}, \bibinfo{person}{Hui Zeng}, {and}
  \bibinfo{person}{Yibo Fan}.} \bibinfo{year}{2022}\natexlab{}.
\newblock \showarticletitle{APMC: Adjacent Pixels Based Measurement Coding
  System for Compressively Sensed Images}.
\newblock \bibinfo{journal}{\emph{IEEE Transactions on Multimedia}}
  \bibinfo{volume}{24} (\bibinfo{year}{2022}), \bibinfo{pages}{3558--3569}.
\newblock
\urldef\tempurl%
\url{https://doi.org/10.1109/TMM.2021.3102394}
\showDOI{\tempurl}


\bibitem[Wang et~al\mbox{.}(2023)]%
        {10124848}
\bibfield{author}{\bibinfo{person}{Huake Wang}, \bibinfo{person}{Ziang Li},
  {and} \bibinfo{person}{Xingsong Hou}.} \bibinfo{year}{2023}\natexlab{}.
\newblock \showarticletitle{Versatile Denoising-Based Approximate Message
  Passing for Compressive Sensing}.
\newblock \bibinfo{journal}{\emph{IEEE Transactions on Image Processing}}
  \bibinfo{volume}{32} (\bibinfo{year}{2023}), \bibinfo{pages}{2761--2775}.
\newblock



\bibitem[Wiegand et~al\mbox{.}(2003)]%
        {Wiegand2003Overview}
\bibfield{author}{\bibinfo{person}{Thomas Wiegand}, \bibinfo{person}{Gary~J
  Sullivan}, \bibinfo{person}{Gisle Bj{\o}ntegaard}, {and}
  \bibinfo{person}{Ajay Luthra}.} \bibinfo{year}{2003}\natexlab{}.
\newblock \showarticletitle{Overview of the {H}. 264/{AVC} video coding
  standard}.
\newblock \bibinfo{journal}{\emph{IEEE Transactions on Circuits and Systems for
  Video Technology}} \bibinfo{volume}{13}, \bibinfo{number}{7}
  (\bibinfo{year}{2003}), \bibinfo{pages}{560--576}.
\newblock


\bibitem[Wright et~al\mbox{.}(2009)]%
        {wright2009sparse}
\bibfield{author}{\bibinfo{person}{Stephen~J Wright}, \bibinfo{person}{Robert~D
  Nowak}, {and} \bibinfo{person}{M{\'a}rio~AT Figueiredo}.}
  \bibinfo{year}{2009}\natexlab{}.
\newblock \showarticletitle{Sparse reconstruction by separable approximation}.
\newblock \bibinfo{journal}{\emph{IEEE Transactions on Signal Processing}}
  \bibinfo{volume}{57}, \bibinfo{number}{7} (\bibinfo{year}{2009}),
  \bibinfo{pages}{2479--2493}.
\newblock


\bibitem[Wu et~al\mbox{.}(2016)]%
        {wu2016privacy}
\bibfield{author}{\bibinfo{person}{Dapeng Wu}, \bibinfo{person}{Boran Yang},
  \bibinfo{person}{Honggang Wang}, \bibinfo{person}{Chonggang Wang}, {and}
  \bibinfo{person}{Ruyan Wang}.} \bibinfo{year}{2016}\natexlab{}.
\newblock \showarticletitle{Privacy-preserving multimedia big data aggregation
  in large-scale wireless sensor networks}.
\newblock \bibinfo{journal}{\emph{ACM Transactions on Multimedia Computing,
  Communications, and Applications (TOMM)}} \bibinfo{volume}{12},
  \bibinfo{number}{4s} (\bibinfo{year}{2016}), \bibinfo{pages}{1--19}.
\newblock


\bibitem[Xu et~al\mbox{.}(2014)]%
        {xu2014compressive}
\bibfield{author}{\bibinfo{person}{Hongwei Xu}, \bibinfo{person}{Ning Fu},
  \bibinfo{person}{Liyan Qiao}, \bibinfo{person}{Wei Yu}, {and}
  \bibinfo{person}{Xiyuan Peng}.} \bibinfo{year}{2014}\natexlab{}.
\newblock \showarticletitle{Compressive blind mixing matrix estimation of audio
  signals}.
\newblock \bibinfo{journal}{\emph{IEEE Transactions on Instrumentation and
  Measurement}} \bibinfo{volume}{63}, \bibinfo{number}{5}
  (\bibinfo{year}{2014}), \bibinfo{pages}{1253--1261}.
\newblock


\bibitem[Xu et~al\mbox{.}(2018)]%
        {xu2018lapran}
\bibfield{author}{\bibinfo{person}{Kai Xu}, \bibinfo{person}{Zhikang Zhang},
  {and} \bibinfo{person}{Fengbo Ren}.} \bibinfo{year}{2018}\natexlab{}.
\newblock \showarticletitle{LAPRAN: A Scalable Laplacian Pyramid Reconsructive
  Adversarial Network for Flexible Compressive Sensing Reconstruction}.
\newblock \bibinfo{journal}{\emph{Springer European Conference on Computer
  Vision (ECCV)}} (\bibinfo{year}{2018}), \bibinfo{pages}{491--507}.
\newblock


\bibitem[Yan et~al\mbox{.}(2014)]%
        {yan2014shrinkage}
\bibfield{author}{\bibinfo{person}{Wenjie Yan}, \bibinfo{person}{Qiang Wang},
  {and} \bibinfo{person}{Yi Shen}.} \bibinfo{year}{2014}\natexlab{}.
\newblock \showarticletitle{Shrinkage-based alternating projection algorithm
  for efficient measurement matrix construction in compressive sensing}.
\newblock \bibinfo{journal}{\emph{IEEE Transactions on Instrumentation and
  Measurement}} \bibinfo{volume}{63}, \bibinfo{number}{5}
  (\bibinfo{year}{2014}), \bibinfo{pages}{1073--1084}.
\newblock


\bibitem[Yang et~al\mbox{.}(2021)]%
        {10.1145/3447431}
\bibfield{author}{\bibinfo{person}{Peihao Yang}, \bibinfo{person}{Linghe Kong},
  \bibinfo{person}{Meikang Qiu}, \bibinfo{person}{Xue Liu}, {and}
  \bibinfo{person}{Guihai Chen}.} \bibinfo{year}{2021}\natexlab{}.
\newblock \showarticletitle{Compressed Imaging Reconstruction with Sparse
  Random Projection}.
\newblock \bibinfo{journal}{\emph{ACM Trans. Multimedia Comput. Commun. Appl.}}
  \bibinfo{volume}{17}, \bibinfo{number}{1}, Article \bibinfo{articleno}{26}
  (\bibinfo{date}{apr} \bibinfo{year}{2021}), \bibinfo{numpages}{25}~pages.
\newblock
\showISSN{1551-6857}
\urldef\tempurl%
\url{https://doi.org/10.1145/3447431}
\showDOI{\tempurl}


\bibitem[Yang et~al\mbox{.}(2013)]%
        {Yang2013Variational}
\bibfield{author}{\bibinfo{person}{Zai Yang}, \bibinfo{person}{Lihua Xie},
  {and} \bibinfo{person}{Cishen Zhang}.} \bibinfo{year}{2013}\natexlab{}.
\newblock \showarticletitle{Variational Bayesian Algorithm for Quantized
  Compressed Sensing}.
\newblock \bibinfo{journal}{\emph{IEEE Transactions on Signal Processing}}
  \bibinfo{volume}{61}, \bibinfo{number}{11} (\bibinfo{year}{2013}),
  \bibinfo{pages}{2815--2824}.
\newblock


\bibitem[Yao et~al\mbox{.}(2017)]%
        {Yao2017DR2}
\bibfield{author}{\bibinfo{person}{Hantao Yao}, \bibinfo{person}{Feng Dai},
  \bibinfo{person}{Dongming Zhang}, \bibinfo{person}{Yike Ma},
  \bibinfo{person}{Shiliang Zhang}, \bibinfo{person}{Yongdong Zhang}, {and}
  \bibinfo{person}{Qi Tian}.} \bibinfo{year}{2017}\natexlab{}.
\newblock \showarticletitle{DR2-Net: Deep Residual Reconstruction Network for
  Image Compressive Sensing}.
\newblock \bibinfo{journal}{\emph{IEEE Conference on Computer Vision and
  Pattern Recognition (CVPR)}} (\bibinfo{year}{2017}).
\newblock


\bibitem[You et~al\mbox{.}(2021)]%
        {9467810}
\bibfield{author}{\bibinfo{person}{Di You}, \bibinfo{person}{Jian Zhang},
  \bibinfo{person}{Jingfen Xie}, \bibinfo{person}{Bin Chen}, {and}
  \bibinfo{person}{Siwei Ma}.} \bibinfo{year}{2021}\natexlab{}.
\newblock \showarticletitle{COAST: COntrollable Arbitrary-Sampling NeTwork for
  Compressive Sensing}.
\newblock \bibinfo{journal}{\emph{IEEE Transactions on Image Processing}}
  \bibinfo{volume}{30} (\bibinfo{year}{2021}), \bibinfo{pages}{6066--6080}.
\newblock
\urldef\tempurl%
\url{https://doi.org/10.1109/TIP.2021.3091834}
\showDOI{\tempurl}


\bibitem[Yuan and Haimi-Cohen(2020)]%
        {8962013}
\bibfield{author}{\bibinfo{person}{Xin Yuan} {and} \bibinfo{person}{Raziel
  Haimi-Cohen}.} \bibinfo{year}{2020}\natexlab{}.
\newblock \showarticletitle{Image Compression Based on Compressive Sensing:
  End-to-End Comparison With {JPEG}}.
\newblock \bibinfo{journal}{\emph{IEEE Transactions on Multimedia}}
  \bibinfo{volume}{22}, \bibinfo{number}{11} (\bibinfo{year}{2020}),
  \bibinfo{pages}{2889--2904}.
\newblock
\urldef\tempurl%
\url{https://doi.org/10.1109/TMM.2020.2967646}
\showDOI{\tempurl}


\bibitem[Zhang and Ghanem(2018)]%
        {zhang2018ista}
\bibfield{author}{\bibinfo{person}{Jian Zhang} {and} \bibinfo{person}{Bernard
  Ghanem}.} \bibinfo{year}{2018}\natexlab{}.
\newblock \showarticletitle{ISTA-Net: Interpretable Optimization-Inspired Deep
  Network for Image Compressive Sensing}.
\newblock \bibinfo{journal}{\emph{IEEE Conference on Computer Vision and
  Pattern Recognition (CVPR)}} (\bibinfo{year}{2018}),
  \bibinfo{pages}{1828--1837}.
\newblock


\bibitem[Zhang et~al\mbox{.}(2020)]%
        {9019857}
\bibfield{author}{\bibinfo{person}{Jian Zhang}, \bibinfo{person}{Chen Zhao},
  {and} \bibinfo{person}{Wen Gao}.} \bibinfo{year}{2020}\natexlab{}.
\newblock \showarticletitle{Optimization-Inspired Compact Deep Compressive
  Sensing}.
\newblock \bibinfo{journal}{\emph{IEEE Journal of Selected Topics in Signal
  Processing}} \bibinfo{volume}{14}, \bibinfo{number}{4}
  (\bibinfo{year}{2020}), \bibinfo{pages}{765--774}.
\newblock
\urldef\tempurl%
\url{https://doi.org/10.1109/JSTSP.2020.2977507}
\showDOI{\tempurl}


\bibitem[Zhang et~al\mbox{.}(2014)]%
        {zhang2014group}
\bibfield{author}{\bibinfo{person}{Jian Zhang}, \bibinfo{person}{Debin Zhao},
  {and} \bibinfo{person}{Wen Gao}.} \bibinfo{year}{2014}\natexlab{}.
\newblock \showarticletitle{Group-based Sparse Representation for Image
  Restoration}.
\newblock \bibinfo{journal}{\emph{IEEE Transactions on Image Processing (TIP)}}
  \bibinfo{volume}{23}, \bibinfo{number}{8} (\bibinfo{year}{2014}),
  \bibinfo{pages}{3336--3351}.
\newblock


\bibitem[Zhang et~al\mbox{.}(2013)]%
        {Zhang2013Spatially}
\bibfield{author}{\bibinfo{person}{Jian Zhang}, \bibinfo{person}{Debin Zhao},
  {and} \bibinfo{person}{Feng Jiang}.} \bibinfo{year}{2013}\natexlab{}.
\newblock \showarticletitle{Spatially Directional Predictive Coding for
  Block-based Compressive Sensing of Natural Images}.
\newblock \bibinfo{journal}{\emph{IEEE International Conference on Image
  Processing (ICIP)}} (\bibinfo{year}{2013}), \bibinfo{pages}{1021--1025}.
\newblock


\bibitem[Zhang et~al\mbox{.}(2012)]%
        {zhang2012compressed}
\bibfield{author}{\bibinfo{person}{Jian Zhang}, \bibinfo{person}{Debin Zhao},
  \bibinfo{person}{Chen Zhao}, \bibinfo{person}{Ruiqin Xiong},
  \bibinfo{person}{Siwei Ma}, {and} \bibinfo{person}{Wen Gao}.}
  \bibinfo{year}{2012}\natexlab{}.
\newblock \showarticletitle{Compressed sensing recovery via collaborative
  sparsity}. In \bibinfo{booktitle}{\emph{Data Compression Conference (DCC),
  2012}}. IEEE, \bibinfo{pages}{287--296}.
\newblock


\bibitem[Zhang et~al\mbox{.}(2021)]%
        {9298950}
\bibfield{author}{\bibinfo{person}{Zhonghao Zhang}, \bibinfo{person}{Yipeng
  Liu}, \bibinfo{person}{Jiani Liu}, \bibinfo{person}{Fei Wen}, {and}
  \bibinfo{person}{Ce Zhu}.} \bibinfo{year}{2021}\natexlab{}.
\newblock \showarticletitle{AMP-Net: Denoising-Based Deep Unfolding for
  Compressive Image Sensing}.
\newblock \bibinfo{journal}{\emph{IEEE Transactions on Image Processing}}
  \bibinfo{volume}{30} (\bibinfo{year}{2021}), \bibinfo{pages}{1487--1500}.
\newblock
\urldef\tempurl%
\url{https://doi.org/10.1109/TIP.2020.3044472}
\showDOI{\tempurl}


\bibitem[Zhao et~al\mbox{.}(2016)]%
        {zhao2016nonconvex}
\bibfield{author}{\bibinfo{person}{Chen Zhao}, \bibinfo{person}{Jian Zhang},
  \bibinfo{person}{Siwei Ma}, {and} \bibinfo{person}{Wen Gao}.}
  \bibinfo{year}{2016}\natexlab{}.
\newblock \showarticletitle{Nonconvex lp nuclear norm based admm framework for
  compressed sensing}. In \bibinfo{booktitle}{\emph{2016 Data Compression
  Conference (DCC)}}. IEEE, \bibinfo{pages}{161--170}.
\newblock


\bibitem[{Zhao} et~al\mbox{.}(2019)]%
        {8445655}
\bibfield{author}{\bibinfo{person}{L. {Zhao}}, \bibinfo{person}{H. {Bai}},
  \bibinfo{person}{A. {Wang}}, {and} \bibinfo{person}{Y. {Zhao}}.}
  \bibinfo{year}{2019}\natexlab{}.
\newblock \showarticletitle{Multiple Description Convolutional Neural Networks
  for Image Compression}.
\newblock \bibinfo{journal}{\emph{IEEE Transactions on Circuits and Systems for
  Video Technology}} \bibinfo{volume}{29}, \bibinfo{number}{8}
  (\bibinfo{year}{2019}), \bibinfo{pages}{2494--2508}.
\newblock
\urldef\tempurl%
\url{https://doi.org/10.1109/TCSVT.2018.2867067}
\showDOI{\tempurl}


\bibitem[{Zhu} et~al\mbox{.}(2019)]%
        {8629276}
\bibfield{author}{\bibinfo{person}{S. {Zhu}}, \bibinfo{person}{C. {Cui}},
  \bibinfo{person}{R. {Xiong}}, \bibinfo{person}{Y. {Guo}}, {and}
  \bibinfo{person}{B. {Zeng}}.} \bibinfo{year}{2019}\natexlab{}.
\newblock \showarticletitle{Efficient Chroma Sub-Sampling and Luma Modification
  for Color Image Compression}.
\newblock \bibinfo{journal}{\emph{IEEE Transactions on Circuits and Systems for
  Video Technology}} \bibinfo{volume}{29}, \bibinfo{number}{5}
  (\bibinfo{year}{2019}), \bibinfo{pages}{1559--1563}.
\newblock
\urldef\tempurl%
\url{https://doi.org/10.1109/TCSVT.2019.2895840}
\showDOI{\tempurl}


\end{thebibliography}

\end{document}